\documentclass[reprint,superscriptaddress]{revtex4-1}

\usepackage{graphicx}
\usepackage{amsmath}
\usepackage{color,soul}
\begin{document}
\newcommand{\Tau}{\mathrm{T}}
\title{Cross-correlation signal processing for axion and WISP dark matter searches}

\author{Ben T. McAllister}
\email{ben.mcallister@uwa.edu.au}
\author{Stephen R. Parker}
\affiliation{ARC Centre of Excellence for Engineered Quantum Systems, School of Physics, The University of Western Australia, Crawley 6009, Australia}
\author{Eugene N. Ivanov}
\affiliation{School of Physics, The University of Western Australia, Crawley 6009, Australia}
\author{Michael E. Tobar}
\email{michael.tobar@uwa.edu.au}
\affiliation{ARC Centre of Excellence for Engineered Quantum Systems, School of Physics, The University of Western Australia, Crawley 6009, Australia}

\date{\today}

\begin{abstract}
The search for dark matter is of fundamental importance to our understanding of the universe. Weakly-Interacting Slim Particles (WISPs) such as axions and hidden sector photons (HSPs) are well motivated candidates for the dark matter. Some of the most sensitive and mature experiments to detect WISPs rely on microwave cavities, and the detection of weak photon signals. It is often suggested to power combine multiple cavities, which creates a host of technical concerns. We outline a scheme based on cross-correlation for effectively power combining cavities and increasing the signal-to-noise ratio of a candidate WISP signal.
\end{abstract}


\maketitle

\section{Introduction}
Many problems in particle physics can be solved via the introduction of weakly-interacting slim (sub-eV) particles (WISPs)~\cite{wisps}. For example, one of the most elegant solutions to the strong CP problem in quantum chromodynamics (QCD) relies on the introduction of a new spin zero, neutral boson, a WISP known as the axion~\cite{PQ1977,Weinberg1978,Wilczek1978,Kim2010}. Axions, and other WISPs such as Hidden Sector Photons (HSPs)~\cite{sme1,sme2,holdem,paraphoton} can be formulated as compelling dark matter candidates~\cite{Sikivie1983,Preskill1983,Dine1983,Sikivie1983b,nelson2011,arias2012}. The axion in particular is an appealing dark matter candidate, which arises from a separate area of physics, and exhibits the desired properties.\\
The most mature detection techniques for WISPs typically rely on WISP-to-photon couplings. These couplings are typically among the most well formulated and least model-dependent WISP interactions. Different WISPs couple to photons via different mechanisms. For example, the axion couples to two photons, and axion-photon conversion can be induced when a magnetic field sources a virtual photon (inverse Primakoff effect~\cite{Sikivie83haloscope,Sikivie1985}), while Hidden Sector Photons undergo spontaneous kinetic mixing analogous to neutrino flavour oscillations. However, in both cases detection reduces to the detection of a photon with a frequency corresponding to the mass of the WISP in question. There are large parameter spaces associated with the various WISP dark matter candidates, as most of the properties of these particles are only weakly-constrained by cosmological observations and previous experiments. However, we can focus on some expected ranges for masses and coupling strengths based on different theoretical models. Most WISP dark matter models predict masses in the $\mu$eV to meV range, corresponding to photons in the microwave and millimetre-wave regimes~\cite{Kim2010,wisps}.\\
When it comes to detecting dark matter WISPs, potentially the most sensitive and mature technique is known as a haloscope~\cite{Sikivie83haloscope,hagmann1990,Bradley2003}, so called as it searches for particles in the galactic halo. The details of such a search differ slightly depending on whether you intend to search for axions or HSPs, but in both cases the WISP dark matter flux converts into photons under certain conditions~\cite{McAllisterPRL}, and the goal is to detect this small photon signal above the thermal and technical noise in the readout of a cryogenic resonant cavity, which is present so that the signal may be resonantly enhanced.\\
Another sensitive WISP detection technique is the so called Light Shining Through a Wall experiment~\cite{Jaeckel08,Graham2014}. In such an experiment a single cavity is pumped with photons on resonance and the correct conditions are supplied such that the WISP of interest (be it an axion or a HSP) will be generated from this cavity via the inverse of the coupling mechanism used in a haloscope. A second cavity is present to detect these WISPs via a conversion process, similar to a haloscope, the only difference being that the origin of the WISP in such an experiment is not galactic halo dark matter, but rather the first cavity. Such experiments require less assumptions about the nature of these WISP particles (they do not require the particles to constitute halo dark matter), but are inherently less sensitive as they require two WISP-photon conversions.\\
Several microwave haloscope~\cite{ADMXaxions2010,ADMX2011,CAPP,CAPPToroid,HAYSTAC,ORGAN} and LSW~\cite{povey2010,ADMX2010,betz2013,parker2013b,Yale2014} experiments have been conducted to date, with no reports of any signals congruent with WISP detection. The majority of these searches have been conducted below 10~GHz due to a multitude of technical issues and limitations. Despite this, the higher frequency parameter space is theoretically well motivated and has even offered up some hints of possible CDM WISPs~\cite{Beck2013}. Some recent work suggests that 50-200~$\mu$eV, corresponding to 12.5-50 GHz is a promising region for axion searches~\cite{SMASH}. Much recent work has focused on ways to access this higher mass axion regime~\cite{DielectricRing,AxionArray,CAPPPIZZA}. Techniques such as the one detailed here will become increasingly useful in the push to larger numbers of cavities required for sensitive high frequency WISP searches, such as the planned ORGAN Experiment~\cite{ORGAN}.\\
The remainder of this work focuses on an outline of a cross-correlation technique for combining multiple resonators in cavity WISP searches. We wish to emphasize that the proposed technique is applicable to a number of different types of experiments~\cite{povey2011,parker2013,altmethods2013,Seviour2014,Sikivie2014a,BudkerPRX,altmethods2014,Sikivie2014}.
\section{WISP Detection \label{sec:haloscope}}
\begin{figure}[t]
\centering
\includegraphics[width=0.8\columnwidth]{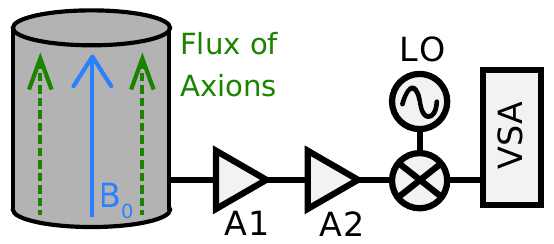}
\caption{Schematic of a WISP cavity receiver, in this case the prototypical axion haloscope. A1 represents the first-stage amplifier, while A2 covers all further stages of amplification. LO is the Local Oscillator used to down-convert the cavity spectrum for sampling on a Fast Fourier Transform Vector Signal Analyzer (VSA), or some appropriate digitizer.}
\label{fig:haloscope}
\end{figure}

Figure~\ref{fig:haloscope} is a diagram of a haloscope designed to detect dark matter axions. In such an experiment a static, external magnetic field is applied to provide a source of virtual photons, and if dark matter axions are present in the cavity, a small number should convert to photons such that
\begin{equation}
\text{h}f_a \approx \text{m}_a\text{c}^2,
\end{equation}
with some line-broadening due to velocity dispersion. The generated photons have a mode structure such that the electric field follows the applied static magnetic field. If the cavity resonance is tuned to overlap with the frequency of these photons, and if the cavity mode overlaps with the mode structure of the generated photon signal, a build up of power will occur in the cavity, given by
\begin{equation}
P_{a}\propto\left(\frac{g_{\gamma}\alpha}{\pi f_{a}}\right)^{2}\frac{\rho_{a}}{m_{a}}VB_{0}^{2}C\text{min}\left(Q_{L},Q_{a}\right), \label{eq:paxion3}
\end{equation}
\\
where $g_{\gamma}$ is a model-dependent parameter of $\mathcal{O}$(1)~\cite{K79,SVZ80,DFS81,Kim2010}, $\alpha$ is the fine structure constant and $f_{a}$ is the Peccei-Quinn symmetry-breaking scale~\cite{Kim2010}. This energy scale dictates the mass of the axion and the strength of its coupling to standard model particles; this is the parameter that haloscopes ultimately aim to measure or bound. The strength of the expected axion power signal also depends on the local density of axion dark matter, $\rho_{a}$. Cosmological observations provide an estimate for the local dark matter density (0.35$^{+0.08}_{-0.07}$~GeV/cm$^{3}$~\cite{cdmdensity2014}), however it is important to note that to date axion haloscope searches have typically only excluded CDM axions under the assumption that they constitute all of the local dark matter. It is plausible that dark matter is comprised of more than one particle species and thus the local axion dark matter density is lower than has been assumed in previous haloscope searches. This opens up the possibility of needing to repeat axion searches in already excluded regions of parameter space to check for lower density axion CDM.\\
The remaining factors of eq.~\eqref{eq:paxion3} relate to the properties of the cavity used - $V$ is the volume, $Q_{L}$ is the loaded quality factor and $C$ is a form factor~\cite{McAllisterPRL} describing the relative overlap of the axion-induced electromagnetic field and the electromagnetic field of the chosen cavity resonance, which is typically the TM$_{010}$ mode for haloscopes ($C\sim$0.69 for an empty cylindrical cavity). The so-called axion quality factor, $Q_{a}$, describes the expected structure of a virialized CDM galactic halo axion signal, where considerations of dispersion lead to a value of $\sim10^{6}$. Nonvirialized axion CDM would have a narrower linewidth (higher $Q_{a}$)~\cite{ADMX2011}, approaching that of a quasi-monochromatic signal.\\
In a haloscope, or similar cavity receiver, the ultimate goal is to resolve an additional photon signal (due to the WISP of interest) above the noise background. The degree of resolution of such a signal directly relates to the confidence in the observed results. For WISP searches, signal-to-noise ratios of 5 or more are typically desired. The primary noise contributions for cavity receivers come from the intrinsic thermal noise of the resonator, and the technical noise of the first stage amplifier, with later amplifier noise contributions effectively suppressed by the gain of this amplifier~\cite{Bradley2003}. We model the amplifier as a pure white noise source followed by a pure gain, and assume critical coupling of the cavity receiver to the readout chain.
We can express the Power Spectral Density (PSD, in units Watts/Hz) of cavity thermal fluctuations at the input of the amplifier as
\begin{align}
\text{S}_{\text{cav}} = k_{B}\text{T}_{0}\mathrm{T(j\omega)}, 
\label{eq:cavnoise}
\end{align}
where T$_{0}$ is the physical temperature of the cavity and $\mathrm{T(j\omega)}$ is the Lorentzian transmission coefficient of the cavity. Similarly the PSD of broadband amplifier white noise referred to the amplifier input is  given by
\begin{equation}
\text{S}_{\text{amp}} = k_{B}\text{T}_{\text{eff}}, \label{eq:ampnoise}
\end{equation}
where T$_{\text{eff}}$ is the sum of the effective noise temperature of the first stage amplifier and its physical temperature. This is assuming that the first stage amplifier and data acquisition system are operating at a frequency above a few hundred kHz, outside the $\frac{1}{f}$ noise regime.
For the following discussion, we shall define the SNR of a WISP cavity receiver as
\begin{equation}
\text{SNR}=\frac{\text{S}^{\text{max}}_{x}-\left<\text{S}_{x}\right>}{\sigma_{\text{s}_{x}}},
\label{eq:SNRdef}
\end{equation}
where
\begin{equation}
\text{S}^{\text{max}}_{x}=\text{S}_{\text{sig}}+\text{S}_{x}.
\end{equation}
Here S$_{x}$, $\left<\text{S}_{x}\right>$ and $\sigma_{\text{s}_{x}}$ are the value of the background noise PSD at the frequency of interest (the sum of (3) and (4)) and the mean and standard deviation of the PSD in the absence of the WISP signal. $\text{S}_{\text{sig}}$ is the value of the PSD of the WISP signal at the frequency of interest, given by
\begin{equation}
\text{S}_{\text{sig}}\approx\frac{P_\text{sig}}{\Delta f_W},
\end{equation}
Where $\Delta f_W$ is the linewidth of the WISP signal. The value of the noise PSD, $\text{S}_x$, at the frequency of interest will be a random variable with a mean $\left<\text{S}_x\right>  = \left<\text{S}_\text{cav}\right> + \left<\text{S}_\text{amp}\right>$ with some random fluctuations, so
\begin{equation}
\text{S}_x=\left<\text{S}_x\right>+ k\sigma_{\text{s}_x},
\end{equation}
where k is a random variable which tells us how many standard deviations the value of the noise PSD is away from its mean, in the absence of a WISP signal at the frequency of interest. Seeing as we cannot distinguish between a WISP signal and the random fluctuations around this mean value, and can only refer the power in a given bin to the mean of the background, the \emph{effective} WISP signal in a given bin is given by
\begin{equation}
\text{S}^\text{eff}_\text{sig}=\text{S}_\text{sig}+ k\sigma_{\text{s}_x},
\end{equation}
and equation~\eqref{eq:SNRdef} becomes
\begin{equation}
\text{SNR}=\frac{\text{S}_{\text{sig}}+k\sigma_{\text{s}_x}}{\sigma_{\text{s}_x}}=\frac{\text{S}^{\text{eff}}_{sig}}{\sigma_{\text{s}_x}}.
\end{equation}
\begin{figure}[t]
\centering
\includegraphics[width=0.8\columnwidth]{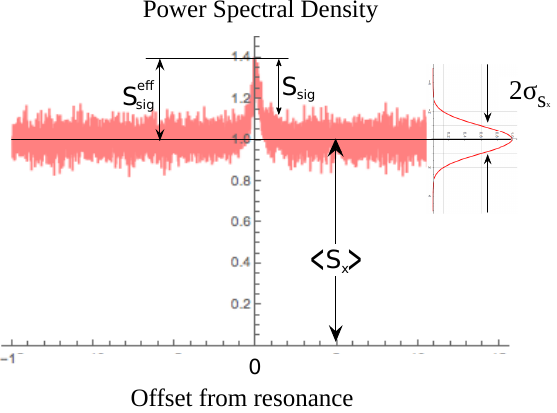}
\caption{Sketch illustrating the signal-to-noise ratio of a signal of interest in a WISP cavity receiver. Several key parameters discussed in the text are made explicit in the figure. The effective signal is the deviation of the PSD from the mean of the WISP signal-free background.}
\label{fig:SNRsketch}
\end{figure}
See fig.~\ref{fig:SNRsketch}, which illustrates many of these concepts.
Now, for random noise averaged $m$ times on a single channel, it is a well known result that
\begin{equation}
\sigma_{\text{s}_x}=\frac{\left<\text{S}_x\right>}{\sqrt{m}}.
\end{equation}
Thus, we finally arrive at
\begin{equation}
SNR=\frac{\text{S}^{\text{eff}}_{sig}}{\left<\text{S}_x\right>}\sqrt{m}.
\label{eq:SNRReal}
\end{equation}
It should be stressed that $\text{S}^{\text{eff}}_{sig}$ is simply a measure of how far the combined PSD is away from the mean at a given frequency. This definition of signal to noise is slightly different to the commonly used Radiometer equation, but it contains the same sources (ie the combined power on resonance from the WISP and the receiver, as well as the background noise from the receiver), and scales by root $m$ as it should. The measure of signal-to-noise ratio in eq.~\eqref{eq:SNRReal} is useful for comparing different cavity detection schemes. With real data, eq.~\eqref{eq:SNRdef} (which simply measures how many standard deviations the power in a given bin is away from the mean) should be employed for analysis. Defined in the manner above, any SNR value is equivalent to the number of standard deviations a signal is away from the mean, ie a SNR of 3 corresponds to a signal 3 sigma away from the mean, and a SNR of 5 would represent the 5 sigma standard threshold for detection employed in particle physics.
\begin{figure}[t!]
\centering
\includegraphics[width=0.3\columnwidth]{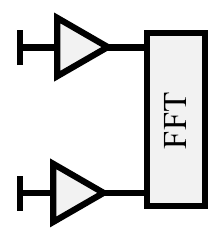}
\caption{Schematic of the measurement scheme used to demonstrate principles of cross-correlation. Both amplifiers were SR560 low-noise pre-amplifier models from Stanford Research Systems with separate 50~$\Omega$ terminations attached to each input. Spectrums were recorded on the FFT with 1601 points in the frequency range 10-100~kHz, well outside of the flicker noise regime of the amplifiers.}
\label{fig:xsnfschem}
\end{figure}
\section{Cross-Correlation}
Cross-correlation measurements~\cite{WallsCC,Rubiola,Rubiola2010} have been used for many years in the characterization of low noise devices such as microwave amplifiers and oscillators. The technique involves computing the cross-correlation function of two signals; any fluctuations that are not correlated between the signals are rejected, what remains is any process that is present in both signals and correlated. Applied correctly this provides a powerful tool to reduce or eliminate the noise contributions of measurement systems. Cross-correlation measurements are practically limited by the level of isolation between the two measurement channels, typically it is possible to achieve rejection of the uncorrelated noise on the order of 20 - 30~dB. This is by no means a fundamental limit, merely from observation and readily achievable levels of isolation in the laboratory with standard components.
\begin{figure}[t]
\centering
\includegraphics[width=0.8\columnwidth]{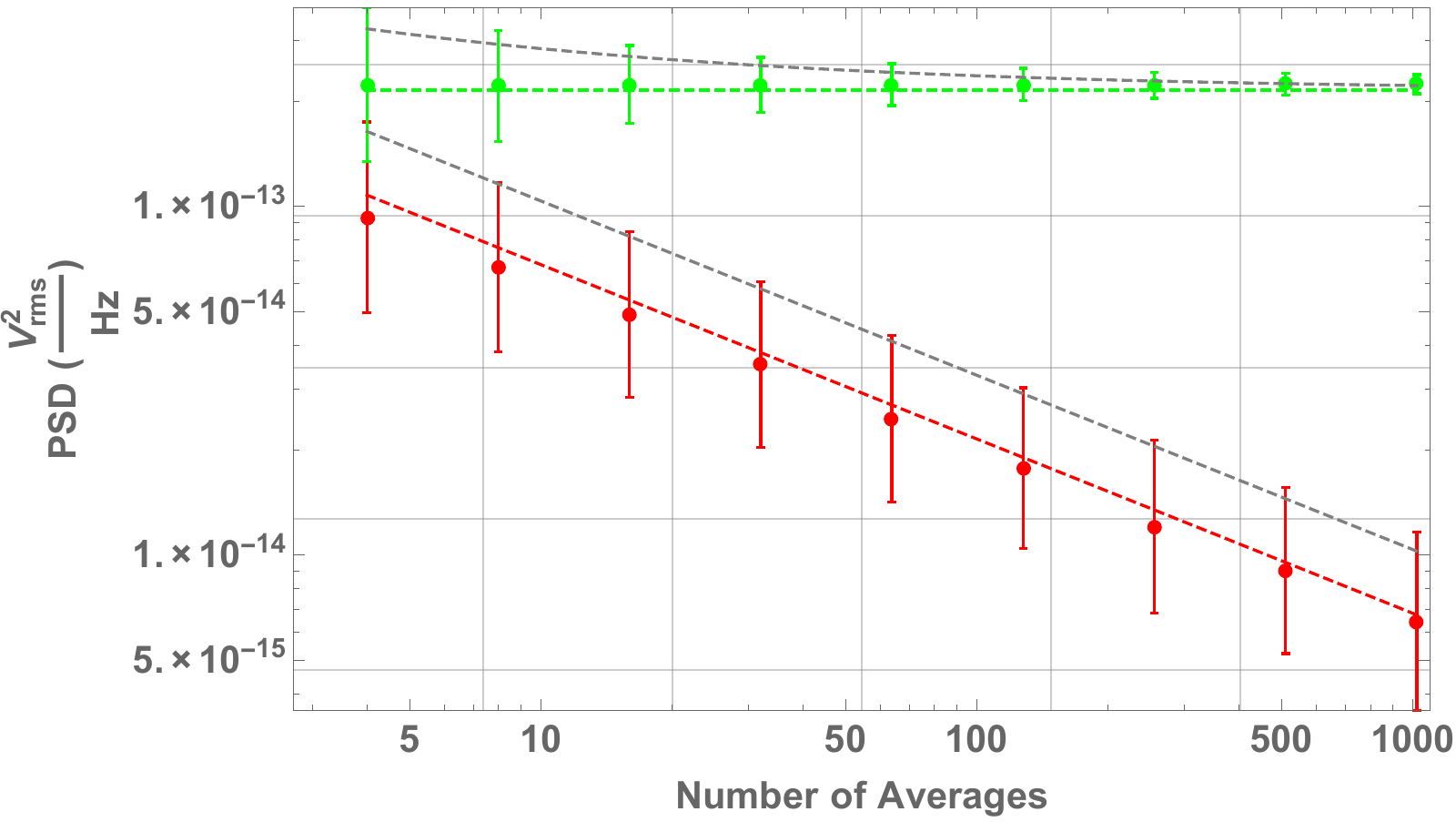}
\caption{Mean of voltage spectra as a function of averages from the measurement scheme presented in fig.~\ref{fig:xsnfschem}. The mean of a single channel spectrum is shown (green circles) with error bars representing 1 standard deviation. A $\sqrt{m}$ dependence is shown (grey dashed line) for the standard deviation. The mean of the cross-spectrum is also shown (red circles) with error bars representing 1 standard deviation. The function $\mu_{1}$/$\sqrt{m}$ is plotted (red dashed line) as discussed in the text.}
\label{fig:xsreject}
\end{figure}
The cross-spectrum of two signals, $a(t)$ and $b(t)$, is defined as the Fourier Transform of the cross-correlation function,
\begin{equation}
S^{ab}\left(f\right)=\mathcal{F}\left\{\lim_{\theta\to\infty}\frac{1}{\theta}\int_{\theta}a(t)b(t-\tau)dt\right\}.
\label{eq:cs}
\end{equation}
However, experimentally we deal with digitzed signals such that $a(t)$ and $b(t)$ have been sampled and now contain $N$ discrete values. The discrete cross-correlation function is
\begin{equation}
\text{K}_{ab}^{\left(i\right)}=\frac{1}{N}\sum\limits_{k=1}^{N}a_{k}b_{k+|i-N|}, \label{eq:dcc}
\end{equation}
where the index $i$ runs from 1 to 2$N$. Hence, the discrete implementation of eq.~\eqref{eq:cs} is
\begin{equation}
S^{ab}\left(f_{i}\right)=\mathcal{DF}\left\{\text{K}_{ab}^{\left(i\right)}\right\}, \label{eq:dxs}
\end{equation}
where $\mathcal{DF}$ is the discrete Fourier transform. The cross-power spectral density (X-PSD) can then be obtained by normalizing eq.~\eqref{eq:dxs} over the resolution bandwidth (i.e. multiplying by $\tau_{\text{meas}}$).\\
We now consider the effect of averaging over $m$ spectra and cross-spectra. During this process if $a(t)$ and $b(t)$ contain any signals that are uncorrelated they will be rejected from the cross-spectrum at a rate proportional to $\sqrt{m}$~\cite{Rubiola,Rubiola2010}, until the limit of isolation between the two channels is reached.\\
For a single channel measurement, the error (standard deviation) is reduced with $\sqrt{m}$ while the mean, $\mu$, remains constant. In the cross-spectrum, the mean is reduced by $\sqrt{m}$ while the error (standard deviation) remains proportionally constant relative to the mean (and thus also decreases by an absolute factor of $\sqrt{m}$). The standard deviation in the cross-spectrum is roughly half of the standard deviation in the single channel, for a given number of averages, $m$.
Once the limit of rejection is reached, the standard deviation of the cross-spectrum starts to reduce relative to the mean. Most WISP searches are unlikely to reach the limit of isolation between measurement channels.\\
We performed a simple measurement to demonstrate the principles of cross-correlation outlined in this section. Figure~\ref{fig:xsnfschem} shows the method used; data was collected for averages ranging from 4 to 1024. The mean of the single channel spectrum is shown (green circles) in fig.~\ref{fig:xsreject} with the error bars representing 1 standard deviation. The $\sqrt{m}$ dependence of the error is shown by the dashed grey lines, while the mean ($\mu_{1}$) stays constant. The mean value of the cross-spectrum (red circles) is also presented along with 1 standard deviation error bars. The function $\mu_{1}$/$\sqrt{m}$ is plotted (red dashed line) and overlaps with the mean of the cross-spectrum, while the standard deviation of the cross-spectrum remains proportionally constant relative to the mean, and offset by a factor of $\sim2$ from the standard deviation of the single channel.
\section{Cross-Correlation WISP Cavity Receiver Measurement Techniques \label{sec:cc}}
\begin{figure}[t]
\centering
\includegraphics[width=0.8\columnwidth]{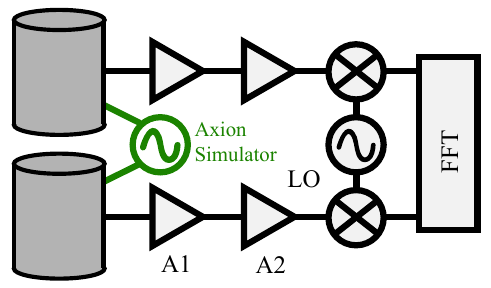}
\caption{Schematic of the multi-cavity cross-correlation measurement technique. The axion simulator (green) is used in sec.~\ref{sec:results} to perform proof-of-concept measurements.}
\label{fig:cc2cav}
\end{figure}
A scheme for applying cross-correlation techniques to WISP searches (in this case, a haloscope) is outlined in fig.~\ref{fig:cc2cav}. In this setup, two separate but identical cavities are each placed inside a strong magnetic field, and read out via independent measurement channels. Each cavity has its own amplification chain. This differs from other multiple cavity haloscope schemes, which typically involve power combining multiple resonators prior to amplification and readout via a single chain~\cite{darin2001,ADMXHF2014}. Seeing as the thermal cavity noise and amplifier technical noise are uncorrelated between the two channels, when the cross-spectrum is computed these noise sources are rejected from the final spectrum. A WISP signal will remain correlated between the two channels provided that the receivers are not separated by a distance greater than the de Broglie wavelength of the particle in question, which for axions is of the order of tens of metres~\cite{carosi2008}. Phase offset between the two channels should not impact the degree of noise rejection, as the noise is all thermal (random) in origin, and completely independent between channels.\\
Now, much like a traditional Wilkinson power summation technique, we increase the experimental complexity. We must now frequency tune two cavities so that the resonances overlap. However, even in the situation of imperfect frequency matching, this technique will still be beneficial.
To optimize the experiment the WISP mass should coincide with the resonant frequencies of the cavities, and if one of the cavities was frequency detuned this can be treated as a reduction in the effective $Q$-factor of the cavity~\cite{parker2013b}, which would reduce our ability to resolve a signal. As long as the cavities are not frequency detuned further apart than the bandwidth of the resonant cavity mode being used then this technique is still applicable, albeit degraded.
In the situation where the two cavities and readout chains are identical, we can scale eq.~\eqref{eq:SNRReal} accordingly. Since for the cross-spectrum
\begin{align*}
\sigma_N&\approx\frac{\left<\text{S}_N\right>}{2\sqrt{m}},
\end{align*}
which is a factor of $\sim2$ less than the spread of the single channel, we can conclude via the same reasoning that brought us to eq.~\eqref{eq:SNRReal} 
\begin{align*}
\text{SNR}&\approx\frac{\text{S}^{eff}_{sig}}{\frac{\left<\text{S}_N\right>}{2\sqrt{m}}}\\
&\approx2\sqrt{m}\frac{\text{S}^{eff}_{sig}}{\left<\text{S}_N\right>}.
\end{align*}
This is a factor of $\sim$2 improvement over~\eqref{eq:SNRReal}. Whilst this technique does not offer an immediately clear advantage over a traditional Wilkinson power summation technique for multiple cavities (yielding the same increase in SNR), it does have some potential benefits which could be exploited.\\
First of all, this technique would be useful for characterizing a candidate WISP signal. It would be possible to increase the distance between two cavities that were being combined in this fashion until the correlated signal was lost. This would give an indication of the coherence length, and hence the de Broglie wavelength of the particle in question. Furthermore, this technique allows the two cavities to be spatially well separated, whereas the Wilkinson technique typically requires that they be co-located to reduce additional losses due to the length of transmission lines from the cavities to the power combiner. Indeed, in the cross-correlation scheme one could acquire the data for each measurement channel simultaneously and independently, and then compute the cross-spectrum in post-processing. Finally, using a cross-correlation scheme, we do not need to concern ourselves with the phase difference between the two channels being combined, as we must in the Wilkinson power combining scheme.\\
Additionally, the cross-correlation scheme presented here can be extrapolated to include $n$ cavities. In a situation where we have $n$ cavities we are able to compute all possible cross-spectra in post-processing. For $n$ cavities this means $n\left(n-1\right)/2$ distinct cross-spectra. If we average these $n\left(n-1\right)/2$ cross-spectra linearly with one another, we can view this as equivalent to taking $\sim n\left(n-1\right)/2$ times as many averages of a single cross-spectrum computed from two cavity and readout spectra. The effective increase by a factor of $\sim n\left(n-1\right)/2$ in the number of averages leads to an effective decrease in the spread of the noise by a factor of $\sim\sqrt{n\left(n-1\right)/2}$. As a result, eq.~\eqref{eq:SNRReal} becomes
\begin{equation}
\text{SNR}\approx2\sqrt{\frac{mn(n-1)}{2}}\frac{\text{S}^{eff}_{sig}}{\left<\text{S}_N\right>}.
\label{eq:nxpsd}
\end{equation}
This is a small improvement over the factor of $n$ that one acquires from Wilkinson power combining $n$ cavities, given by
\begin{equation}
n\sqrt{m}\frac{\text{S}_{eff}^{\text{sig}}}{\left<\text{S}_N\right>}.
\label{eq:v3}
\end{equation}
This improvement may seem counter-intuitive, but when one considers that cross-correlating $n$ cavities over $m$ measurements requires us to take $n\times m$ total measurements, versus just $m$ measurements for the Wilkinson power combined scheme, and that this data is then post-processed, it is perhaps easier to intuit how an improvement can be achieved with a larger number of total measurements. This could result in an increased total measurement time if we were forced to take the measurements one after another, however, provided we have enough measurement devices to sample the $n$ readouts simultaneously, we will not incur a time penalty. We do, however, require $n$ amplifiers, as opposed to just one in the Wilkinson scheme.
\section{Results \label{sec:results}}
\begin{figure}[t!]
\centering
\includegraphics[width=0.8\columnwidth]{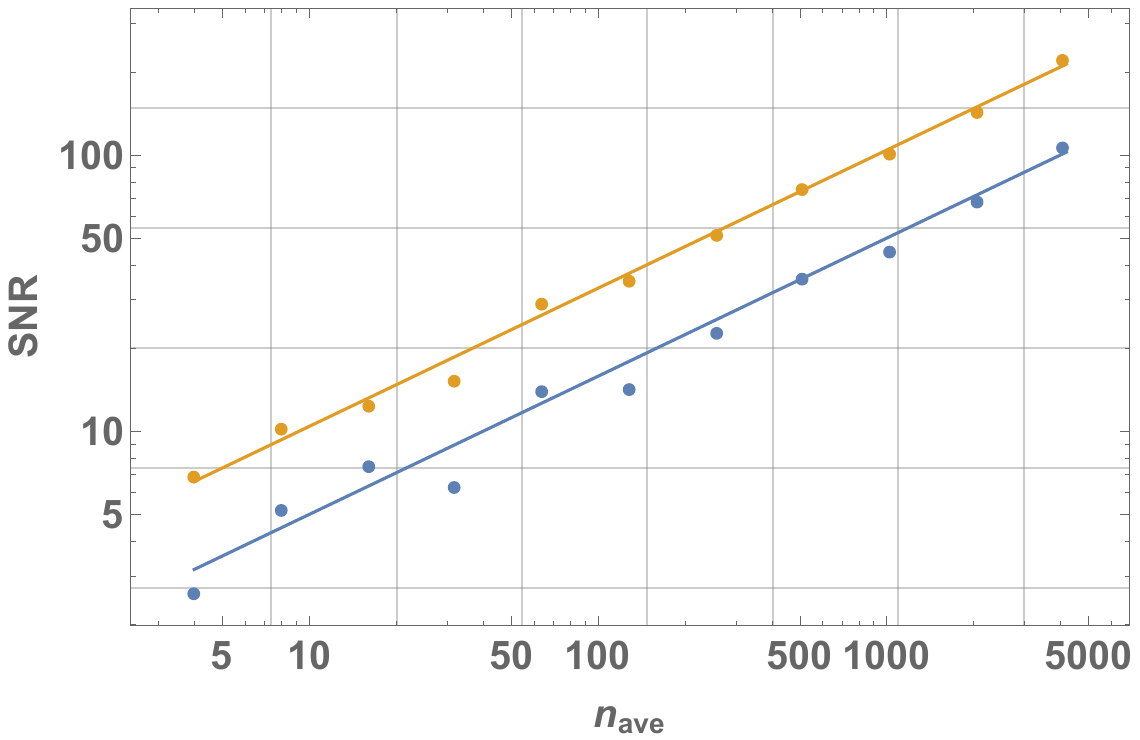}
\caption{Log-log representation of SNR as a function of averages for the simulated axion signal in single channel (blue circles) and X-PSD (yellow circles) obtained via the technique presented in fig.~\ref{fig:cc2cav}. Fits to the data (coloured lines) are presented as discussed in the text.}
\label{fig:xssnr}
\end{figure}

\begin{figure}[t!]
	\includegraphics[width=0.9\columnwidth]{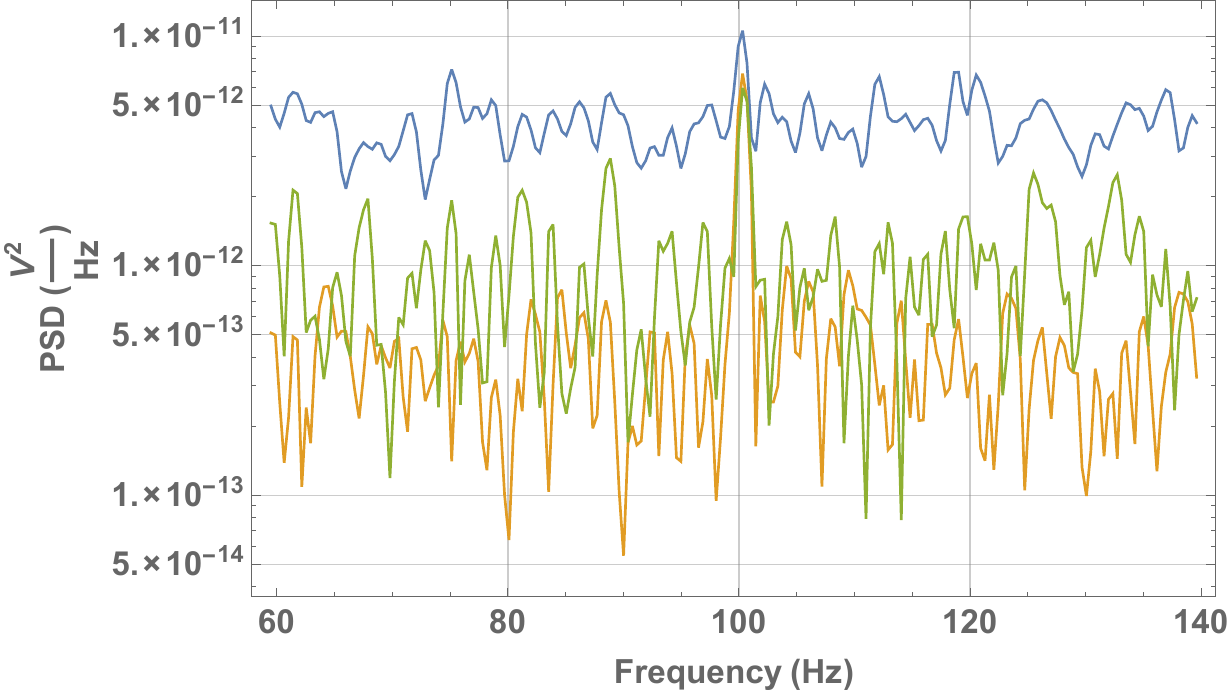}
	\caption{PSD (blue), X-PSD (green), and AX-PSD (yellow) after 16 averages for the four channel cross-correlation scheme described in the text. The broadband noise of the amplifier is seen to decrease for the X-PSD and AX-PSD when compared with the single channel.}
	\label{fig:4chanspec}
\end{figure}

\begin{figure}[h!]
	\centering
	\includegraphics[width=0.8\columnwidth]{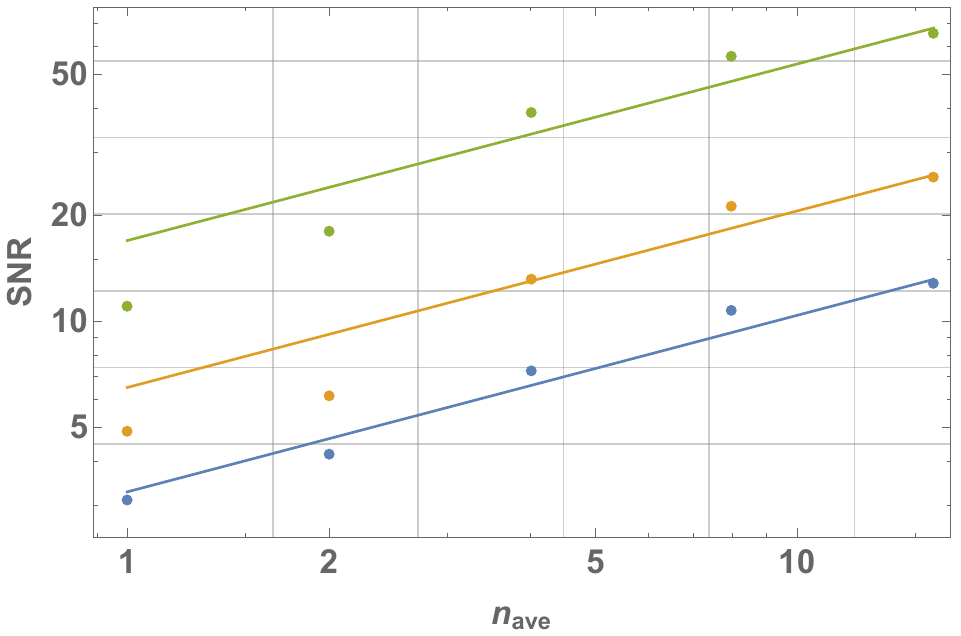}
	\caption{Log-log representation of SNR as a function of averages for the simulated axion signal into a single amplification chain (blue), the average of the SNR for each of the 6 computed X-PSDs (yellow), and the SNR of the averaged X-PSD or AX-PSD (green). Fits to the data (coloured lines) are presented as discussed in the text.}
	\label{fig:4chanSNR}
\end{figure}

Proof-of-concept measurements were made with a two cavity system. A microwave synthesizer was used to generate a simulated WISP signal that falls within the cavity bandwidth. Measurements were conducted at room temperature using commercial sapphire-based ``black box" resonators. As this is not an actual WISP search but rather a test of a measurement technique the mode structure of the cavity resonance is irrelevant. The cavities both had resonance frequencies at $\sim$9~GHz and $Q$ factors of $\sim2\times10^{5}$. The total gain in each channel was $\sim$30~dB. A microwave synthesizer and mixers were employed to mix the channels down to 4~MHz so that the spectra could be recorded on a commercial Vector Signal Analyzer (sometimes known as a Fast Fourier Transform machine (FFT)). Measurements were made from 2 to 4096 averages. With the appropriate equipment such as a FFT or multi-channel digitizer both channels can be sampled simultaneously, and the cross-spectrum is computed in situ, meaning that there is no increase in acquisition time for cross-correlation measurements compared to single channel measurements.
\begin{figure}[t]
\centering
\includegraphics[width=0.8\columnwidth]{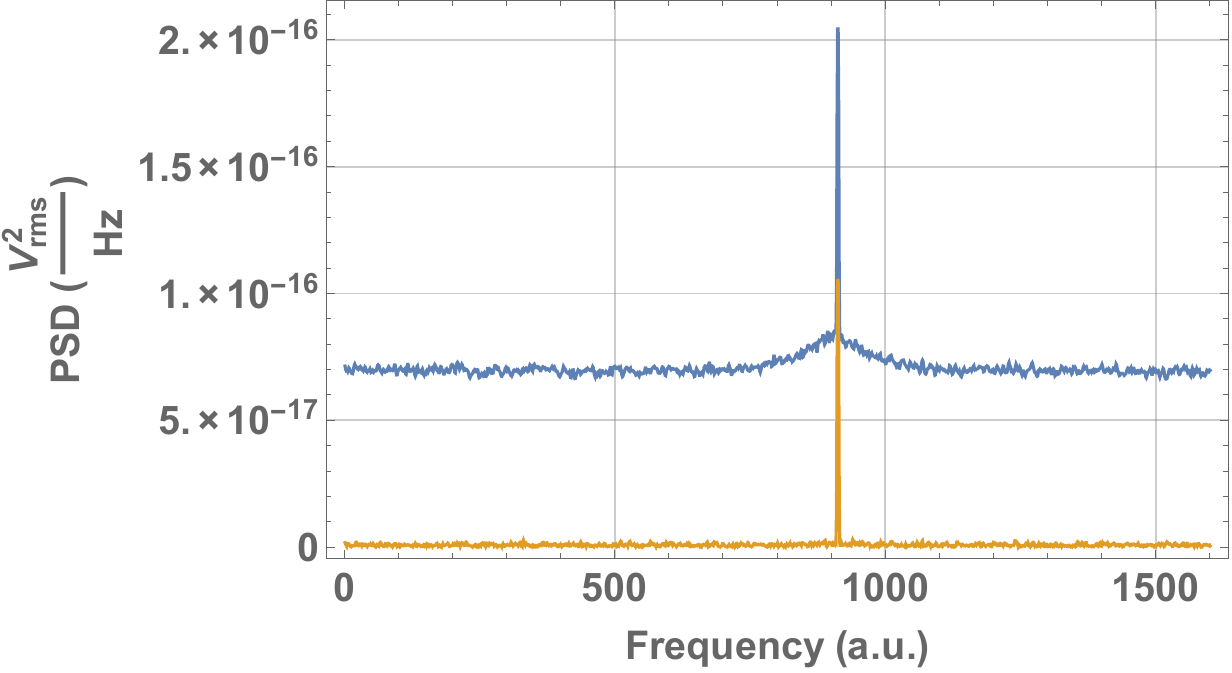}
\caption{A 1600~point PSD for the two cavity cross-correlation technique illustrated in fig.~\ref{fig:cc2cav}. Single channel trace is shown in blue and the X-PSD is shown in yellow. Note that the horizontal axis units are not Hertz, they simply refer to the number assigned to each of the 1600 points in the spectrum.}
\label{fig:spec3}
\end{figure}
Single channel and dual channel results for the two cavity technique (fig.~\ref{fig:cc2cav}) are shown in fig.~\ref{fig:spec3}. Temperature control was used to perform minor frequency tuning to ensure that the resonant frequencies of both cavities were equal. After 4096 averages the single channel trace, representative of a typical WISP cavity receiver, has a SNR of 106$\sigma$ for the WISP signal. In the X-PSD, the first-stage amplifier noise and thermal cavity noise from both channels has been rejected (within the limitations of the instrumentation and the number of averages taken). The simulated WISP signal which is correlated between both cavities remains with an enhanced SNR of 218$\sigma$, which is a factor of $\sim2$ improvement arising from the use of two independent measurement channels. Figure~\ref{fig:xssnr} shows the SNR for a single channel and the X-PSD as a function of the number of averages taken. Fitting to the X-PSD gives the function 3.3$\times\sqrt{m}$ and fitting to a single channel gives 1.6$\times\sqrt{m}$. This demonstrates that the system behaves as outlined in section~\ref{sec:cc} and also shows that the X-PSD conserves relatively weak correlated signals, as the starting SNR for these measurements is less than 1 for the single channel. Averages below 4 have been omitted from the figure as the signal cannot be resolved above the background noise.\\
Another tabletop experiment was performed to verify the expected sensitivity predicted by eq.~\eqref{eq:nxpsd} for cross-correlating combinations of $n$ channels. Four independent amplifiers were injected with a simulated monochromatic WISP signal of -106 dBm at 1200 Hz, the outputs being sampled directly by a 4 channel digitizer. The gain of each amplifier was set to 2 dB. The data was processed to compute PSDs for each channel, as well as all 6 possible combinations of X-PSDs for averages ranging from 1 to 16. Subsequently the averaged cross-power spectral density (AX-PSD) was computed via linearly averaging the 6 individual X-PSDs. Results for the single channel, cross-spectrum and averaged cross-spectrum after 16 averages are shown in fig.~\ref{fig:4chanspec}. After 16 averages the average SNR for the single channel was 12.8$\sigma$. When examining the X-PSDs, the average SNR was 25.4$\sigma$, which is close to the factor of 2 expected. When examining the averaged cross-spectrum the SNR was found to be 65.2$\sigma$, which is slightly better than the factor of 2$\sqrt{6}$ expected from eq.~\eqref{eq:nxpsd}, when compared with the average single channel value. Figure~\ref{fig:4chanSNR} shows the SNR for PSD, X-PSD and AX-PSD as a function of the number of averages taken.\\
Fitting to the AX-PSD gives the function 16.9$\sqrt{m}$, fitting to the X-PSD gives the function 6.48$\sqrt{m}$, and fitting to the single channel gives the function 3.28$\sqrt{m}$. These values show a factor of $\sim2$ between the single channel PSD and the X-PSD as expected, and a factor slightly larger than the expected 2$\sqrt{6}$ between the PSD and the AX-PSD. This demonstrates that at least for the four independent amplifier channels the system performs as expected. It should be noted that this technique was not attempted with cavity structures due to technical limitations (we did not have 4 identical cavity and amplifier readouts on hand). Furthermore it is important to stress that for both the 2 cavity and 4 channel schemes tested here, much care must be taken to ensure that the signals in each channel are nearly identical. Discrepancy between the channels leads to deviation from the predictions outlined in this text.\\
For both measurement techniques the phase offset in each arm, or more importantly the phase difference between the channels, ultimately has minimal impact upon the SNR observed in the X-PSD. Regardless, it would be prudent to fully characterize any system deployed in an actual data-taking experiment.
\section{Conclusion} 
A cross-correlation measurement scheme for cavity-based WISP searches has been proposed and tested. This scheme allows for two cavity outputs to be effectively combined via computation of their cross-power spectral density. This allows for the cavities to be spatially well separated, thereby providing a way to measure the coherence length of any candidate WISP signal. As a further benefit, the relative phase in the two channels has no impact on the combined sensitivity, which simplifies the experiment. Furthermore, such an approach can be scaled up to larger arrays of cavities by computing all the possible combinations of cavity cross-power spectral densities, which results in an improvement in SNR versus what can be achieved by power combining the same number of cavities. We require more total measurements, amplifiers and readout chains, but provided we have the requisite hardware this is not an issue, and we can reach an improved sensitivity in the same amount of time.\\
The authors thank J.M.~Le-Floch for help with the data acquisition software. This work was funded by Australian Research Council grant No. DP160100253 and CE170100009, the Australian Government’s Research Training Program, and the Bruce and Betty Green Foundation.


\begin{thebibliography}{51}%
		\makeatletter
		\providecommand \@ifxundefined [1]{%
			\@ifx{#1\undefined}
		}%
		\providecommand \@ifnum [1]{%
			\ifnum #1\expandafter \@firstoftwo
			\else \expandafter \@secondoftwo
			\fi
		}%
		\providecommand \@ifx [1]{%
			\ifx #1\expandafter \@firstoftwo
			\else \expandafter \@secondoftwo
			\fi
		}%
		\providecommand \natexlab [1]{#1}%
		\providecommand \enquote  [1]{``#1''}%
		\providecommand \bibnamefont  [1]{#1}%
		\providecommand \bibfnamefont [1]{#1}%
		\providecommand \citenamefont [1]{#1}%
		\providecommand \href@noop [0]{\@secondoftwo}%
		\providecommand \href [0]{\begingroup \@sanitize@url \@href}%
		\providecommand \@href[1]{\@@startlink{#1}\@@href}%
		\providecommand \@@href[1]{\endgroup#1\@@endlink}%
		\providecommand \@sanitize@url [0]{\catcode `\\12\catcode `\$12\catcode
			`\&12\catcode `\#12\catcode `\^12\catcode `\_12\catcode `\%12\relax}%
		\providecommand \@@startlink[1]{}%
		\providecommand \@@endlink[0]{}%
		\providecommand \url  [0]{\begingroup\@sanitize@url \@url }%
		\providecommand \@url [1]{\endgroup\@href {#1}{\urlprefix }}%
		\providecommand \urlprefix  [0]{URL }%
		\providecommand \Eprint [0]{\href }%
		\providecommand \doibase [0]{http://dx.doi.org/}%
		\providecommand \selectlanguage [0]{\@gobble}%
		\providecommand \bibinfo  [0]{\@secondoftwo}%
		\providecommand \bibfield  [0]{\@secondoftwo}%
		\providecommand \translation [1]{[#1]}%
		\providecommand \BibitemOpen [0]{}%
		\providecommand \bibitemStop [0]{}%
		\providecommand \bibitemNoStop [0]{.\EOS\space}%
		\providecommand \EOS [0]{\spacefactor3000\relax}%
		\providecommand \BibitemShut  [1]{\csname bibitem#1\endcsname}%
		\let\auto@bib@innerbib\@empty
		\bibitem [{\citenamefont {Jaeckel}\ and\ \citenamefont
			{Ringwald}(2010)}]{wisps}%
		\BibitemOpen
		\bibfield  {author} {\bibinfo {author} {\bibfnamefont {J.}~\bibnamefont
				{Jaeckel}}\ and\ \bibinfo {author} {\bibfnamefont {A.}~\bibnamefont
				{Ringwald}},\ }\href {\doibase 10.1146/annurev.nucl.012809.104433} {\bibfield
			{journal} {\bibinfo  {journal} {Annual Review of Nuclear and Particle
					Science}\ }\textbf {\bibinfo {volume} {60}},\ \bibinfo {pages} {405}
			(\bibinfo {year} {2010})}\BibitemShut {NoStop}%
		\bibitem [{\citenamefont {Peccei}\ and\ \citenamefont {Quinn}(1977)}]{PQ1977}%
		\BibitemOpen
		\bibfield  {author} {\bibinfo {author} {\bibfnamefont {R.~D.}\ \bibnamefont
				{Peccei}}\ and\ \bibinfo {author} {\bibfnamefont {H.~R.}\ \bibnamefont
				{Quinn}},\ }\href {\doibase 10.1103/PhysRevLett.38.1440} {\bibfield
			{journal} {\bibinfo  {journal} {Phys. Rev. Lett.}\ }\textbf {\bibinfo
				{volume} {38}},\ \bibinfo {pages} {1440} (\bibinfo {year}
			{1977})}\BibitemShut {NoStop}%
		\bibitem [{\citenamefont {Weinberg}(1978)}]{Weinberg1978}%
		\BibitemOpen
		\bibfield  {author} {\bibinfo {author} {\bibfnamefont {S.}~\bibnamefont
				{Weinberg}},\ }\href {\doibase 10.1103/PhysRevLett.40.223} {\bibfield
			{journal} {\bibinfo  {journal} {Phys. Rev. Lett.}\ }\textbf {\bibinfo
				{volume} {40}},\ \bibinfo {pages} {223} (\bibinfo {year} {1978})}\BibitemShut
		{NoStop}%
		\bibitem [{\citenamefont {Wilczek}(1978)}]{Wilczek1978}%
		\BibitemOpen
		\bibfield  {author} {\bibinfo {author} {\bibfnamefont {F.}~\bibnamefont
				{Wilczek}},\ }\href {\doibase 10.1103/PhysRevLett.40.279} {\bibfield
			{journal} {\bibinfo  {journal} {Phys. Rev. Lett.}\ }\textbf {\bibinfo
				{volume} {40}},\ \bibinfo {pages} {279} (\bibinfo {year} {1978})}\BibitemShut
		{NoStop}%
		\bibitem [{\citenamefont {Kim}\ and\ \citenamefont {Carosi}(2010)}]{Kim2010}%
		\BibitemOpen
		\bibfield  {author} {\bibinfo {author} {\bibfnamefont {J.~E.}\ \bibnamefont
				{Kim}}\ and\ \bibinfo {author} {\bibfnamefont {G.}~\bibnamefont {Carosi}},\
		}\href {\doibase 10.1103/RevModPhys.82.557} {\bibfield  {journal} {\bibinfo
			{journal} {Rev. Mod. Phys.}\ }\textbf {\bibinfo {volume} {82}},\ \bibinfo
		{pages} {557} (\bibinfo {year} {2010})}\BibitemShut {NoStop}%
	
	\bibitem [{\citenamefont {Abel}\ \emph {et~al.}(2008)\citenamefont {Abel},
		\citenamefont {Goodsell}, \citenamefont {Jaeckel}, \citenamefont {Khoze},\
		and\ \citenamefont {Ringwald}}]{sme1}%
	\BibitemOpen
	\bibfield  {author} {\bibinfo {author} {\bibfnamefont {S.}~\bibnamefont
			{Abel}}, \bibinfo {author} {\bibfnamefont {M.}~\bibnamefont {Goodsell}},
		\bibinfo {author} {\bibfnamefont {J.}~\bibnamefont {Jaeckel}}, \bibinfo
		{author} {\bibfnamefont {V.}~\bibnamefont {Khoze}}, \ and\ \bibinfo {author}
		{\bibfnamefont {A.}~\bibnamefont {Ringwald}},\ }\href
	{http://stacks.iop.org/1126-6708/2008/i=07/a=124} {\bibfield  {journal}
		{\bibinfo  {journal} {Journal of High Energy Physics}\ }\textbf {\bibinfo
			{volume} {2008}},\ \bibinfo {pages} {124} (\bibinfo {year}
		{2008})}\BibitemShut {NoStop}%
	\bibitem [{\citenamefont {Goodsell}\ \emph {et~al.}(2009)\citenamefont
		{Goodsell}, \citenamefont {Jaeckel}, \citenamefont {Redondo},\ and\
		\citenamefont {Ringwald}}]{sme2}%
	\BibitemOpen
	\bibfield  {author} {\bibinfo {author} {\bibfnamefont {M.}~\bibnamefont
			{Goodsell}}, \bibinfo {author} {\bibfnamefont {J.}~\bibnamefont {Jaeckel}},
		\bibinfo {author} {\bibfnamefont {J.}~\bibnamefont {Redondo}}, \ and\
		\bibinfo {author} {\bibfnamefont {A.}~\bibnamefont {Ringwald}},\ }\href
	{http://stacks.iop.org/1126-6708/2009/i=11/a=027} {\bibfield  {journal}
		{\bibinfo  {journal} {Journal of High Energy Physics}\ }\textbf {\bibinfo
			{volume} {2009}},\ \bibinfo {pages} {027} (\bibinfo {year}
		{2009})}\BibitemShut {NoStop}%
	\bibitem [{\citenamefont {Holdom}(1986)}]{holdem}%
	\BibitemOpen
	\bibfield  {author} {\bibinfo {author} {\bibfnamefont {B.}~\bibnamefont
			{Holdom}},\ }\href {\doibase 10.1016/0370-2693(86)91377-8} {\bibfield
		{journal} {\bibinfo  {journal} {Physics Letters B}\ }\textbf {\bibinfo
			{volume} {166}},\ \bibinfo {pages} {196 } (\bibinfo {year}
		{1986})}\BibitemShut {NoStop}%
	\bibitem [{\citenamefont {Okun}(1982)}]{paraphoton}%
	\BibitemOpen
	\bibfield  {author} {\bibinfo {author} {\bibfnamefont {L.}~\bibnamefont
			{Okun}},\ }\href@noop {} {\bibfield  {journal} {\bibinfo  {journal}
			{Sov.Phys.JETP}\ }\textbf {\bibinfo {volume} {56}},\ \bibinfo {pages} {502}
		(\bibinfo {year} {1982})}\BibitemShut {NoStop}%
	\bibitem [{\citenamefont {Nelson}\ and\ \citenamefont
		{Scholtz}(2011)}]{nelson2011}%
	\BibitemOpen
	\bibfield  {author} {\bibinfo {author} {\bibfnamefont {A.~E.}\ \bibnamefont
			{Nelson}}\ and\ \bibinfo {author} {\bibfnamefont {J.}~\bibnamefont
			{Scholtz}},\ }\href {\doibase 10.1103/PhysRevD.84.103501} {\bibfield
		{journal} {\bibinfo  {journal} {Phys. Rev. D}\ }\textbf {\bibinfo {volume}
			{84}},\ \bibinfo {pages} {103501} (\bibinfo {year} {2011})}\BibitemShut
	{NoStop}%
	\bibitem [{\citenamefont {Abbott}\ and\ \citenamefont
		{Sikivie}(1983)}]{Sikivie1983}%
	\BibitemOpen
	\bibfield  {author} {\bibinfo {author} {\bibfnamefont {L.}~\bibnamefont
			{Abbott}}\ and\ \bibinfo {author} {\bibfnamefont {P.}~\bibnamefont
			{Sikivie}},\ }\href {\doibase http://dx.doi.org/10.1016/0370-2693(83)90638-X}
	{\bibfield  {journal} {\bibinfo  {journal} {Physics Letters B}\ }\textbf
		{\bibinfo {volume} {120}},\ \bibinfo {pages} {133 } (\bibinfo {year}
		{1983})}\BibitemShut {NoStop}%
	\bibitem [{\citenamefont {Preskill}\ \emph {et~al.}(1983)\citenamefont
		{Preskill}, \citenamefont {Wise},\ and\ \citenamefont
		{Wilczek}}]{Preskill1983}%
	\BibitemOpen
	\bibfield  {author} {\bibinfo {author} {\bibfnamefont {J.}~\bibnamefont
			{Preskill}}, \bibinfo {author} {\bibfnamefont {M.~B.}\ \bibnamefont {Wise}},
		\ and\ \bibinfo {author} {\bibfnamefont {F.}~\bibnamefont {Wilczek}},\ }\href
	{\doibase http://dx.doi.org/10.1016/0370-2693(83)90637-8} {\bibfield
		{journal} {\bibinfo  {journal} {Physics Letters B}\ }\textbf {\bibinfo
			{volume} {120}},\ \bibinfo {pages} {127 } (\bibinfo {year}
		{1983})}\BibitemShut {NoStop}%
	\bibitem [{\citenamefont {Dine}\ and\ \citenamefont
		{Fischler}(1983)}]{Dine1983}%
	\BibitemOpen
	\bibfield  {author} {\bibinfo {author} {\bibfnamefont {M.}~\bibnamefont
			{Dine}}\ and\ \bibinfo {author} {\bibfnamefont {W.}~\bibnamefont
			{Fischler}},\ }\href {\doibase
		http://dx.doi.org/10.1016/0370-2693(83)90639-1} {\bibfield  {journal}
		{\bibinfo  {journal} {Physics Letters B}\ }\textbf {\bibinfo {volume}
			{120}},\ \bibinfo {pages} {137 } (\bibinfo {year} {1983})}\BibitemShut
	{NoStop}%
	\bibitem [{\citenamefont {Ipser}\ and\ \citenamefont
		{Sikivie}(1983)}]{Sikivie1983b}%
	\BibitemOpen
	\bibfield  {author} {\bibinfo {author} {\bibfnamefont {J.}~\bibnamefont
			{Ipser}}\ and\ \bibinfo {author} {\bibfnamefont {P.}~\bibnamefont
			{Sikivie}},\ }\href {\doibase 10.1103/PhysRevLett.50.925} {\bibfield
		{journal} {\bibinfo  {journal} {Phys. Rev. Lett.}\ }\textbf {\bibinfo
			{volume} {50}},\ \bibinfo {pages} {925} (\bibinfo {year} {1983})}\BibitemShut
	{NoStop}%
	\bibitem [{\citenamefont {Arias}\ \emph {et~al.}(2012)\citenamefont {Arias},
		\citenamefont {Cadamuro}, \citenamefont {Goodsell}, \citenamefont {Jaeckel},
		\citenamefont {Redondo},\ and\ \citenamefont {Ringwald}}]{arias2012}%
	\BibitemOpen
	\bibfield  {author} {\bibinfo {author} {\bibfnamefont {P.}~\bibnamefont
			{Arias}}, \bibinfo {author} {\bibfnamefont {D.}~\bibnamefont {Cadamuro}},
		\bibinfo {author} {\bibfnamefont {M.}~\bibnamefont {Goodsell}}, \bibinfo
		{author} {\bibfnamefont {J.}~\bibnamefont {Jaeckel}}, \bibinfo {author}
		{\bibfnamefont {J.}~\bibnamefont {Redondo}}, \ and\ \bibinfo {author}
		{\bibfnamefont {A.}~\bibnamefont {Ringwald}},\ }\href
	{http://stacks.iop.org/1475-7516/2012/i=06/a=013} {\bibfield  {journal}
		{\bibinfo  {journal} {Journal of Cosmology and Astroparticle Physics}\
		}\textbf {\bibinfo {volume} {2012}},\ \bibinfo {pages} {013} (\bibinfo {year}
		{2012})}\BibitemShut {NoStop}%
	\bibitem [{\citenamefont {Sikivie}(1983)}]{Sikivie83haloscope}%
	\BibitemOpen
	\bibfield  {author} {\bibinfo {author} {\bibfnamefont {P.}~\bibnamefont
			{Sikivie}},\ }\href {\doibase 10.1103/PhysRevLett.51.1415} {\bibfield
		{journal} {\bibinfo  {journal} {Phys. Rev. Lett.}\ }\textbf {\bibinfo
			{volume} {51}},\ \bibinfo {pages} {1415} (\bibinfo {year}
		{1983})}\BibitemShut {NoStop}%
	\bibitem [{\citenamefont {Sikivie}(1985)}]{Sikivie1985}%
	\BibitemOpen
	\bibfield  {author} {\bibinfo {author} {\bibfnamefont {P.}~\bibnamefont
			{Sikivie}},\ }\href {\doibase 10.1103/PhysRevD.32.2988} {\bibfield  {journal}
		{\bibinfo  {journal} {Phys. Rev. D}\ }\textbf {\bibinfo {volume} {32}},\
		\bibinfo {pages} {2988} (\bibinfo {year} {1985})}\BibitemShut {NoStop}%
	\bibitem [{\citenamefont {Hagmann}\ \emph {et~al.}(1990)\citenamefont
		{Hagmann}, \citenamefont {Sikivie}, \citenamefont {Sullivan}, \citenamefont
		{Tanner},\ and\ \citenamefont {Cho}}]{hagmann1990}%
	\BibitemOpen
	\bibfield  {author} {\bibinfo {author} {\bibfnamefont {C.}~\bibnamefont
			{Hagmann}}, \bibinfo {author} {\bibfnamefont {P.}~\bibnamefont {Sikivie}},
		\bibinfo {author} {\bibfnamefont {N.}~\bibnamefont {Sullivan}}, \bibinfo
		{author} {\bibfnamefont {D.~B.}\ \bibnamefont {Tanner}}, \ and\ \bibinfo
		{author} {\bibfnamefont {S.-I.}\ \bibnamefont {Cho}},\ }\href {\doibase
		10.1063/1.1141427} {\bibfield  {journal} {\bibinfo  {journal} {Review of
				Scientific Instruments}\ }\textbf {\bibinfo {volume} {61}},\ \bibinfo {pages}
		{1076} (\bibinfo {year} {1990})}\BibitemShut {NoStop}%
	\bibitem [{\citenamefont {Bradley}\ \emph {et~al.}(2003)\citenamefont
		{Bradley}, \citenamefont {Clarke}, \citenamefont {Kinion}, \citenamefont
		{Rosenberg}, \citenamefont {van Bibber}, \citenamefont {Matsuki},
		\citenamefont {M\"uck},\ and\ \citenamefont {Sikivie}}]{Bradley2003}%
	\BibitemOpen
	\bibfield  {author} {\bibinfo {author} {\bibfnamefont {R.}~\bibnamefont
			{Bradley}}, \bibinfo {author} {\bibfnamefont {J.}~\bibnamefont {Clarke}},
		\bibinfo {author} {\bibfnamefont {D.}~\bibnamefont {Kinion}}, \bibinfo
		{author} {\bibfnamefont {L.~J.}\ \bibnamefont {Rosenberg}}, \bibinfo {author}
		{\bibfnamefont {K.}~\bibnamefont {van Bibber}}, \bibinfo {author}
		{\bibfnamefont {S.}~\bibnamefont {Matsuki}}, \bibinfo {author} {\bibfnamefont
			{M.}~\bibnamefont {M\"uck}}, \ and\ \bibinfo {author} {\bibfnamefont
			{P.}~\bibnamefont {Sikivie}},\ }\href {\doibase 10.1103/RevModPhys.75.777}
	{\bibfield  {journal} {\bibinfo  {journal} {Rev. Mod. Phys.}\ }\textbf
		{\bibinfo {volume} {75}},\ \bibinfo {pages} {777} (\bibinfo {year}
		{2003})}\BibitemShut {NoStop}%
		\bibitem [{\citenamefont {McAllister}\ \emph
  {et~al.}(2016{\natexlab{b}})\citenamefont {McAllister}, \citenamefont
  {Parker},\ and\ \citenamefont {Tobar}}]{McAllisterPRL}%
  \BibitemOpen
  \bibfield  {author} {\bibinfo {author} {\bibfnamefont {B.~T.}\ \bibnamefont
  {McAllister}}, \bibinfo {author} {\bibfnamefont {S.~R.}\ \bibnamefont
  {Parker}}, \ and\ \bibinfo {author} {\bibfnamefont {M.~E.}\ \bibnamefont
  {Tobar}},\ }\href {\doibase 10.1103/PhysRevLett.117.159901,
  10.1103/PhysRevLett.116.161804} {\bibfield  {journal} {\bibinfo  {journal}
  {Phys. Rev. Lett.}\ }\textbf {\bibinfo {volume} {116}},\ \bibinfo {pages}
  {161804} (\bibinfo {year} {2016}{\natexlab{b}})},\ \bibinfo {note} {[Erratum:
  Phys. Rev. Lett.117,no.15,159901(2016)]}, \BibitemShut
  {NoStop}%
	\bibitem [{\citenamefont {Jaeckel}\ and\ \citenamefont
		{Ringwald}(2008)}]{Jaeckel08}%
	\BibitemOpen
	\bibfield  {author} {\bibinfo {author} {\bibfnamefont {J.}~\bibnamefont
			{Jaeckel}}\ and\ \bibinfo {author} {\bibfnamefont {A.}~\bibnamefont
			{Ringwald}},\ }\href {\doibase 10.1016/j.physletb.2007.11.071} {\bibfield
		{journal} {\bibinfo  {journal} {Physics Letters B}\ }\textbf {\bibinfo
			{volume} {659}},\ \bibinfo {pages} {509 } (\bibinfo {year}
		{2008})}\BibitemShut {NoStop}%
	\bibitem [{\citenamefont {Graham}\ \emph {et~al.}(2014)\citenamefont {Graham},
		\citenamefont {Mardon}, \citenamefont {Rajendran},\ and\ \citenamefont
		{Zhao}}]{Graham2014}%
	\BibitemOpen
	\bibfield  {author} {\bibinfo {author} {\bibfnamefont {P.~W.}\ \bibnamefont
			{Graham}}, \bibinfo {author} {\bibfnamefont {J.}~\bibnamefont {Mardon}},
		\bibinfo {author} {\bibfnamefont {S.}~\bibnamefont {Rajendran}}, \ and\
		\bibinfo {author} {\bibfnamefont {Y.}~\bibnamefont {Zhao}},\ }\href {\doibase
		10.1103/PhysRevD.90.075017} {\bibfield  {journal} {\bibinfo  {journal} {Phys.
				Rev. D}\ }\textbf {\bibinfo {volume} {90}},\ \bibinfo {pages} {075017}
		(\bibinfo {year} {2014})}\BibitemShut {NoStop}%
	\bibitem [{\citenamefont {Asztalos}\ \emph {et~al.}(2010)\citenamefont
		{Asztalos}, \citenamefont {Carosi}, \citenamefont {Hagmann}, \citenamefont
		{Kinion}, \citenamefont {van Bibber}, \citenamefont {Hotz}, \citenamefont
		{Rosenberg}, \citenamefont {Rybka}, \citenamefont {Hoskins}, \citenamefont
		{Hwang}, \citenamefont {Sikivie}, \citenamefont {Tanner}, \citenamefont
		{Bradley},\ and\ \citenamefont {Clarke}}]{ADMXaxions2010}%
	\BibitemOpen
	\bibfield  {author} {\bibinfo {author} {\bibfnamefont {S.~J.}\ \bibnamefont
			{Asztalos}}, \bibinfo {author} {\bibfnamefont {G.}~\bibnamefont {Carosi}},
		\bibinfo {author} {\bibfnamefont {C.}~\bibnamefont {Hagmann}}, \bibinfo
		{author} {\bibfnamefont {D.}~\bibnamefont {Kinion}}, \bibinfo {author}
		{\bibfnamefont {K.}~\bibnamefont {van Bibber}}, \bibinfo {author}
		{\bibfnamefont {M.}~\bibnamefont {Hotz}}, \bibinfo {author} {\bibfnamefont
			{L.~J.}\ \bibnamefont {Rosenberg}}, \bibinfo {author} {\bibfnamefont
			{G.}~\bibnamefont {Rybka}}, \bibinfo {author} {\bibfnamefont
			{J.}~\bibnamefont {Hoskins}}, \bibinfo {author} {\bibfnamefont
			{J.}~\bibnamefont {Hwang}}, \bibinfo {author} {\bibfnamefont
			{P.}~\bibnamefont {Sikivie}}, \bibinfo {author} {\bibfnamefont {D.~B.}\
			\bibnamefont {Tanner}}, \bibinfo {author} {\bibfnamefont {R.}~\bibnamefont
			{Bradley}}, \ and\ \bibinfo {author} {\bibfnamefont {J.}~\bibnamefont
			{Clarke}},\ }\href {\doibase 10.1103/PhysRevLett.104.041301} {\bibfield
		{journal} {\bibinfo  {journal} {Phys. Rev. Lett.}\ }\textbf {\bibinfo
			{volume} {104}},\ \bibinfo {pages} {041301} (\bibinfo {year}
		{2010})}\BibitemShut {NoStop}%
	\bibitem [{\citenamefont {Hoskins}\ \emph {et~al.}(2011)\citenamefont
		{Hoskins}, \citenamefont {Hwang}, \citenamefont {Martin}, \citenamefont
		{Sikivie}, \citenamefont {Sullivan}, \citenamefont {Tanner}, \citenamefont
		{Hotz}, \citenamefont {Rosenberg}, \citenamefont {Rybka}, \citenamefont
		{Wagner}, \citenamefont {Asztalos}, \citenamefont {Carosi}, \citenamefont
		{Hagmann}, \citenamefont {Kinion}, \citenamefont {van Bibber}, \citenamefont
		{Bradley},\ and\ \citenamefont {Clarke}}]{ADMX2011}%
	\BibitemOpen
	\bibfield  {author} {\bibinfo {author} {\bibfnamefont {J.}~\bibnamefont
			{Hoskins}}, \bibinfo {author} {\bibfnamefont {J.}~\bibnamefont {Hwang}},
		\bibinfo {author} {\bibfnamefont {C.}~\bibnamefont {Martin}}, \bibinfo
		{author} {\bibfnamefont {P.}~\bibnamefont {Sikivie}}, \bibinfo {author}
		{\bibfnamefont {N.~S.}\ \bibnamefont {Sullivan}}, \bibinfo {author}
		{\bibfnamefont {D.~B.}\ \bibnamefont {Tanner}}, \bibinfo {author}
		{\bibfnamefont {M.}~\bibnamefont {Hotz}}, \bibinfo {author} {\bibfnamefont
			{L.~J.}\ \bibnamefont {Rosenberg}}, \bibinfo {author} {\bibfnamefont
			{G.}~\bibnamefont {Rybka}}, \bibinfo {author} {\bibfnamefont
			{A.}~\bibnamefont {Wagner}}, \bibinfo {author} {\bibfnamefont {S.~J.}\
			\bibnamefont {Asztalos}}, \bibinfo {author} {\bibfnamefont {G.}~\bibnamefont
			{Carosi}}, \bibinfo {author} {\bibfnamefont {C.}~\bibnamefont {Hagmann}},
		\bibinfo {author} {\bibfnamefont {D.}~\bibnamefont {Kinion}}, \bibinfo
		{author} {\bibfnamefont {K.}~\bibnamefont {van Bibber}}, \bibinfo {author}
		{\bibfnamefont {R.}~\bibnamefont {Bradley}}, \ and\ \bibinfo {author}
		{\bibfnamefont {J.}~\bibnamefont {Clarke}},\ }\href {\doibase
		10.1103/PhysRevD.84.121302} {\bibfield  {journal} {\bibinfo  {journal} {Phys.
				Rev. D}\ }\textbf {\bibinfo {volume} {84}},\ \bibinfo {pages} {121302}
		(\bibinfo {year} {2011})}\BibitemShut {NoStop}%
\bibitem [{\citenamefont {Chung}(2016)}]{CAPP}%
  \BibitemOpen
  \bibfield  {author} {\bibinfo {author} {\bibfnamefont {W.}~\bibnamefont
  {Chung}},\ }\bibfield  {booktitle} {\emph {\bibinfo {booktitle}
  {{Proceedings, 15th Hellenic School and Workshops on Elementary Particle
  Physics and Gravity (CORFU2015): Corfu, Greece, September 1-25, 2015}}},\
  }\href@noop {} {\bibfield  {journal} {\bibinfo  {journal} {PoS}\ }\textbf
  {\bibinfo {volume} {CORFU2015}},\ \bibinfo {pages} {047} (\bibinfo {year}
  {2016})}\BibitemShut {NoStop}%
\bibitem [{\citenamefont {Choi}\ \emph {et~al.}(2017)\citenamefont {Choi},
  \citenamefont {Themann}, \citenamefont {Lee}, \citenamefont {Ko},\ and\
  \citenamefont {Semertzidis}}]{CAPPToroid}%
  \BibitemOpen
  \bibfield  {author} {\bibinfo {author} {\bibfnamefont {J.}~\bibnamefont
  {Choi}}, \bibinfo {author} {\bibfnamefont {H.}~\bibnamefont {Themann}},
  \bibinfo {author} {\bibfnamefont {M.~J.}\ \bibnamefont {Lee}}, \bibinfo
  {author} {\bibfnamefont {B.~R.}\ \bibnamefont {Ko}}, \ and\ \bibinfo {author}
  {\bibfnamefont {Y.~K.}\ \bibnamefont {Semertzidis}},\ }\href {\doibase
  10.1103/PhysRevD.96.061102} {\bibfield  {journal} {\bibinfo  {journal} {Phys.
  Rev. D}\ }\textbf {\bibinfo {volume} {96}},\ \bibinfo {pages} {061102}
  (\bibinfo {year} {2017})}\BibitemShut {NoStop}%
\bibitem{HAYSTAC} 
  B.~M.~Brubaker {\it et al.},
  Phys.\ Rev.\ Lett.\  {\bf 118}, no. 6, 061302 (2017)
  doi:10.1103/PhysRevLett.118.061302
  [arXiv:1610.02580 [astro-ph.CO]].

	
	
	\bibitem [{\citenamefont {McAllister}\ \emph {et~al.}(2017)\citenamefont
  {McAllister}, \citenamefont {Flower}, \citenamefont {Ivanov}, \citenamefont
  {Goryachev}, \citenamefont {Bourhill},\ and\ \citenamefont {Tobar}}]{ORGAN}%
  \BibitemOpen
  \bibfield  {author} {\bibinfo {author} {\bibfnamefont {B.~T.}\ \bibnamefont
  {McAllister}}, \bibinfo {author} {\bibfnamefont {G.}~\bibnamefont {Flower}},
  \bibinfo {author} {\bibfnamefont {E.~N.}\ \bibnamefont {Ivanov}}, \bibinfo
  {author} {\bibfnamefont {M.}~\bibnamefont {Goryachev}}, \bibinfo {author}
  {\bibfnamefont {J.}~\bibnamefont {Bourhill}}, \ and\ \bibinfo {author}
  {\bibfnamefont {M.~E.}\ \bibnamefont {Tobar}},\ }\href {\doibase
  https://doi.org/10.1016/j.dark.2017.09.010} {\bibfield  {journal} {\bibinfo
  {journal} {Physics of the Dark Universe}\ }\textbf {\bibinfo {volume} {18}},\
  \bibinfo {pages} {67 } (\bibinfo {year} {2017})}\BibitemShut {NoStop}%
	\bibitem [{\citenamefont {Povey}\ \emph {et~al.}(2010)\citenamefont {Povey},
		\citenamefont {Hartnett},\ and\ \citenamefont {Tobar}}]{povey2010}%
	\BibitemOpen
	\bibfield  {author} {\bibinfo {author} {\bibfnamefont {R.~G.}\ \bibnamefont
			{Povey}}, \bibinfo {author} {\bibfnamefont {J.~G.}\ \bibnamefont {Hartnett}},
		\ and\ \bibinfo {author} {\bibfnamefont {M.~E.}\ \bibnamefont {Tobar}},\
	}\href {\doibase 10.1103/PhysRevD.82.052003} {\bibfield  {journal} {\bibinfo
		{journal} {Phys. Rev. D}\ }\textbf {\bibinfo {volume} {82}},\ \bibinfo
	{pages} {052003} (\bibinfo {year} {2010})}\BibitemShut {NoStop}%
\bibitem [{\citenamefont {Wagner}\ \emph {et~al.}(2010)\citenamefont {Wagner},
	\citenamefont {Rybka}, \citenamefont {Hotz}, \citenamefont {Rosenberg},
	\citenamefont {Asztalos}, \citenamefont {Carosi}, \citenamefont {Hagmann},
	\citenamefont {Kinion}, \citenamefont {van Bibber}, \citenamefont {Hoskins},
	\citenamefont {Martin}, \citenamefont {Sikivie}, \citenamefont {Tanner},
	\citenamefont {Bradley},\ and\ \citenamefont {Clarke}}]{ADMX2010}%
\BibitemOpen
\bibfield  {author} {\bibinfo {author} {\bibfnamefont {A.}~\bibnamefont
		{Wagner}}, \bibinfo {author} {\bibfnamefont {G.}~\bibnamefont {Rybka}},
	\bibinfo {author} {\bibfnamefont {M.}~\bibnamefont {Hotz}}, \bibinfo {author}
	{\bibfnamefont {L.~J.}\ \bibnamefont {Rosenberg}}, \bibinfo {author}
	{\bibfnamefont {S.~J.}\ \bibnamefont {Asztalos}}, \bibinfo {author}
	{\bibfnamefont {G.}~\bibnamefont {Carosi}}, \bibinfo {author} {\bibfnamefont
		{C.}~\bibnamefont {Hagmann}}, \bibinfo {author} {\bibfnamefont
		{D.}~\bibnamefont {Kinion}}, \bibinfo {author} {\bibfnamefont
		{K.}~\bibnamefont {van Bibber}}, \bibinfo {author} {\bibfnamefont
		{J.}~\bibnamefont {Hoskins}}, \bibinfo {author} {\bibfnamefont
		{C.}~\bibnamefont {Martin}}, \bibinfo {author} {\bibfnamefont
		{P.}~\bibnamefont {Sikivie}}, \bibinfo {author} {\bibfnamefont {D.~B.}\
		\bibnamefont {Tanner}}, \bibinfo {author} {\bibfnamefont {R.}~\bibnamefont
		{Bradley}}, \ and\ \bibinfo {author} {\bibfnamefont {J.}~\bibnamefont
		{Clarke}},\ }\href {\doibase 10.1103/PhysRevLett.105.171801} {\bibfield
	{journal} {\bibinfo  {journal} {Phys. Rev. Lett.}\ }\textbf {\bibinfo
		{volume} {105}},\ \bibinfo {pages} {171801} (\bibinfo {year}
	{2010})}\BibitemShut {NoStop}%
\bibitem [{\citenamefont {Betz}\ \emph {et~al.}(2013)\citenamefont {Betz},
	\citenamefont {Caspers}, \citenamefont {Gasior}, \citenamefont {Thumm},\ and\
	\citenamefont {Rieger}}]{betz2013}%
\BibitemOpen
\bibfield  {author} {\bibinfo {author} {\bibfnamefont {M.}~\bibnamefont
		{Betz}}, \bibinfo {author} {\bibfnamefont {F.}~\bibnamefont {Caspers}},
	\bibinfo {author} {\bibfnamefont {M.}~\bibnamefont {Gasior}}, \bibinfo
	{author} {\bibfnamefont {M.}~\bibnamefont {Thumm}}, \ and\ \bibinfo {author}
	{\bibfnamefont {S.~W.}\ \bibnamefont {Rieger}},\ }\href {\doibase
	10.1103/PhysRevD.88.075014} {\bibfield  {journal} {\bibinfo  {journal} {Phys.
			Rev. D}\ }\textbf {\bibinfo {volume} {88}},\ \bibinfo {pages} {075014}
	(\bibinfo {year} {2013})}\BibitemShut {NoStop}%
\bibitem [{\citenamefont {Parker}\ \emph
	{et~al.}(2013{\natexlab{a}})\citenamefont {Parker}, \citenamefont {Hartnett},
	\citenamefont {Povey},\ and\ \citenamefont {Tobar}}]{parker2013b}%
\BibitemOpen
\bibfield  {author} {\bibinfo {author} {\bibfnamefont {S.~R.}\ \bibnamefont
		{Parker}}, \bibinfo {author} {\bibfnamefont {J.~G.}\ \bibnamefont
		{Hartnett}}, \bibinfo {author} {\bibfnamefont {R.~G.}\ \bibnamefont {Povey}},
	\ and\ \bibinfo {author} {\bibfnamefont {M.~E.}\ \bibnamefont {Tobar}},\
}\href {\doibase 10.1103/PhysRevD.88.112004} {\bibfield  {journal} {\bibinfo
	{journal} {Phys. Rev. D}\ }\textbf {\bibinfo {volume} {88}},\ \bibinfo
{pages} {112004} (\bibinfo {year} {2013}{\natexlab{a}})}\BibitemShut
{NoStop}%
\bibitem [{\citenamefont {Slocum}\ \emph {et~al.}(2015)\citenamefont {Slocum},
	\citenamefont {Baker}, \citenamefont {Hirshfield}, \citenamefont {Jiang},
	\citenamefont {Malagon}, \citenamefont {Martin}, \citenamefont
	{Shchelkunov},\ and\ \citenamefont {Szymkowiak}}]{Yale2014}%
\BibitemOpen
\bibfield  {author} {\bibinfo {author} {\bibfnamefont {P.}~\bibnamefont
		{Slocum}}, \bibinfo {author} {\bibfnamefont {O.}~\bibnamefont {Baker}},
	\bibinfo {author} {\bibfnamefont {J.}~\bibnamefont {Hirshfield}}, \bibinfo
	{author} {\bibfnamefont {Y.}~\bibnamefont {Jiang}}, \bibinfo {author}
	{\bibfnamefont {A.}~\bibnamefont {Malagon}}, \bibinfo {author} {\bibfnamefont
		{A.}~\bibnamefont {Martin}}, \bibinfo {author} {\bibfnamefont
		{S.}~\bibnamefont {Shchelkunov}}, \ and\ \bibinfo {author} {\bibfnamefont
		{A.}~\bibnamefont {Szymkowiak}},\ }\href {\doibase
	http://dx.doi.org/10.1016/j.nima.2014.10.013} {\bibfield  {journal} {\bibinfo
		{journal} {Nuclear Instruments and Methods in Physics Research Section A:
			Accelerators, Spectrometers, Detectors and Associated Equipment}\ }\textbf
	{\bibinfo {volume} {770}},\ \bibinfo {pages} {76 } (\bibinfo {year}
	{2015})}\BibitemShut {NoStop}%
\bibitem [{\citenamefont {Beck}(2013)}]{Beck2013}%
\BibitemOpen
\bibfield  {author} {\bibinfo {author} {\bibfnamefont {C.}~\bibnamefont
		{Beck}},\ }\href {\doibase 10.1103/PhysRevLett.111.231801} {\bibfield
	{journal} {\bibinfo  {journal} {Phys. Rev. Lett.}\ }\textbf {\bibinfo
		{volume} {111}},\ \bibinfo {pages} {231801} (\bibinfo {year}
	{2013})}\BibitemShut {NoStop}%
\bibitem [{\citenamefont {Ballesteros}\ \emph {et~al.}(2017)\citenamefont
	{Ballesteros}, \citenamefont {Redondo}, \citenamefont {Ringwald},\ and\
	\citenamefont {Tamarit}}]{SMASH}%
\BibitemOpen
\bibfield  {author} {\bibinfo {author} {\bibfnamefont {G.}~\bibnamefont
		{Ballesteros}}, \bibinfo {author} {\bibfnamefont {J.}~\bibnamefont
		{Redondo}}, \bibinfo {author} {\bibfnamefont {A.}~\bibnamefont {Ringwald}}, \
	and\ \bibinfo {author} {\bibfnamefont {C.}~\bibnamefont {Tamarit}},\ }\href
{\doibase 10.1103/PhysRevLett.118.071802} {\bibfield  {journal} {\bibinfo
		{journal} {Phys. Rev. Lett.}\ }\textbf {\bibinfo {volume} {118}},\ \bibinfo
	{pages} {071802} (\bibinfo {year} {2017})},\ \Eprint
{http://arxiv.org/abs/1608.05414} {arXiv:1608.05414 [hep-ph]} \BibitemShut
{NoStop}%
\bibitem [{\citenamefont {McAllister}\ \emph
  {et~al.}(2018{\natexlab{a}})\citenamefont {McAllister}, \citenamefont
  {Flower}, \citenamefont {Tobar},\ and\ \citenamefont
  {Tobar}}]{DielectricRing}%
  \BibitemOpen
  \bibfield  {author} {\bibinfo {author} {\bibfnamefont {B.~T.}\ \bibnamefont
  {McAllister}}, \bibinfo {author} {\bibfnamefont {G.}~\bibnamefont {Flower}},
  \bibinfo {author} {\bibfnamefont {L.~E.}\ \bibnamefont {Tobar}}, \ and\
  \bibinfo {author} {\bibfnamefont {M.~E.}\ \bibnamefont {Tobar}},\ }\href
  {\doibase 10.1103/PhysRevApplied.9.014028} {\bibfield  {journal} {\bibinfo
  {journal} {Phys. Rev. Applied}\ }\textbf {\bibinfo {volume} {9}},\ \bibinfo
  {pages} {014028} (\bibinfo {year} {2018}{\natexlab{a}})}\BibitemShut
  {NoStop}%
\bibitem [{\citenamefont {Goryachev}, \citenamefont {McAllister},\ and\
  \citenamefont {Tobar}(2018)}]{AxionArray}%
  \BibitemOpen
  \bibfield  {author} {\bibinfo {author} {\bibfnamefont {M.}~\bibnamefont
  {Goryachev}}, \bibinfo {author} {\bibfnamefont {B.~T.}\ \bibnamefont
  {McAllister}}, \ and\ \bibinfo {author} {\bibfnamefont {M.~E.}\ \bibnamefont
  {Tobar}},\ }\href {\doibase https://doi.org/10.1016/j.physleta.2017.09.016}
  {\bibfield  {journal} {\bibinfo  {journal} {Physics Letters A}\ }\textbf
  {\bibinfo {volume} {382}},\ \bibinfo {pages} {2199 } (\bibinfo {year}
  {2018})},\ \bibinfo {note} {special Issue in memory of Professor V.B.
  Braginsky}\BibitemShut {NoStop}%
\bibitem [{\citenamefont {Jeong}\ \emph
  {et~al.}(2018{\natexlab{a}})\citenamefont {Jeong}, \citenamefont {Youn},
  \citenamefont {Ahn}, \citenamefont {Kim},\ and\ \citenamefont
  {Semertzidis}}]{CAPPPIZZA}%
  \BibitemOpen
  \bibfield  {author} {\bibinfo {author} {\bibfnamefont {J.}~\bibnamefont
  {Jeong}}, \bibinfo {author} {\bibfnamefont {S.}~\bibnamefont {Youn}},
  \bibinfo {author} {\bibfnamefont {S.}~\bibnamefont {Ahn}}, \bibinfo {author}
  {\bibfnamefont {J.~E.}\ \bibnamefont {Kim}}, \ and\ \bibinfo {author}
  {\bibfnamefont {Y.~K.}\ \bibnamefont {Semertzidis}},\ }\href {\doibase
  https://doi.org/10.1016/j.physletb.2017.12.066} {\bibfield  {journal}
  {\bibinfo  {journal} {Physics Letters B}\ }\textbf {\bibinfo {volume}
  {777}},\ \bibinfo {pages} {412 } (\bibinfo {year}
  {2018}{\natexlab{a}})}\BibitemShut {NoStop}%
\bibitem [{\citenamefont {Povey}\ \emph {et~al.}(2011)\citenamefont {Povey},
	\citenamefont {Hartnett},\ and\ \citenamefont {Tobar}}]{povey2011}%
\BibitemOpen
\bibfield  {author} {\bibinfo {author} {\bibfnamefont {R.~G.}\ \bibnamefont
		{Povey}}, \bibinfo {author} {\bibfnamefont {J.~G.}\ \bibnamefont {Hartnett}},
	\ and\ \bibinfo {author} {\bibfnamefont {M.~E.}\ \bibnamefont {Tobar}},\
}\href {\doibase 10.1103/PhysRevD.84.055023} {\bibfield  {journal} {\bibinfo
	{journal} {Phys. Rev. D}\ }\textbf {\bibinfo {volume} {84}},\ \bibinfo
{pages} {055023} (\bibinfo {year} {2011})}\BibitemShut {NoStop}%
\bibitem [{\citenamefont {Parker}\ \emph
	{et~al.}(2013{\natexlab{b}})\citenamefont {Parker}, \citenamefont {Rybka},\
	and\ \citenamefont {Tobar}}]{parker2013}%
\BibitemOpen
\bibfield  {author} {\bibinfo {author} {\bibfnamefont {S.~R.}\ \bibnamefont
		{Parker}}, \bibinfo {author} {\bibfnamefont {G.}~\bibnamefont {Rybka}}, \
	and\ \bibinfo {author} {\bibfnamefont {M.~E.}\ \bibnamefont {Tobar}},\ }\href
{\doibase 10.1103/PhysRevD.87.115008} {\bibfield  {journal} {\bibinfo
		{journal} {Phys. Rev. D}\ }\textbf {\bibinfo {volume} {87}},\ \bibinfo
	{pages} {115008} (\bibinfo {year} {2013}{\natexlab{b}})}\BibitemShut
{NoStop}%
\bibitem [{\citenamefont {Graham}\ and\ \citenamefont
	{Rajendran}(2013)}]{altmethods2013}%
\BibitemOpen
\bibfield  {author} {\bibinfo {author} {\bibfnamefont {P.~W.}\ \bibnamefont
		{Graham}}\ and\ \bibinfo {author} {\bibfnamefont {S.}~\bibnamefont
		{Rajendran}},\ }\href {\doibase 10.1103/PhysRevD.88.035023} {\bibfield
	{journal} {\bibinfo  {journal} {Phys. Rev. D}\ }\textbf {\bibinfo {volume}
		{88}},\ \bibinfo {pages} {035023} (\bibinfo {year} {2013})}\BibitemShut
{NoStop}%
\bibitem [{\citenamefont {Seviour}\ \emph {et~al.}(2014)\citenamefont
	{Seviour}, \citenamefont {Bailey}, \citenamefont {Woollett},\ and\
	\citenamefont {Williams}}]{Seviour2014}%
\BibitemOpen
\bibfield  {author} {\bibinfo {author} {\bibfnamefont {R.}~\bibnamefont
		{Seviour}}, \bibinfo {author} {\bibfnamefont {I.}~\bibnamefont {Bailey}},
	\bibinfo {author} {\bibfnamefont {N.}~\bibnamefont {Woollett}}, \ and\
	\bibinfo {author} {\bibfnamefont {P.}~\bibnamefont {Williams}},\ }\href
{http://stacks.iop.org/0954-3899/41/i=3/a=035005} {\bibfield  {journal}
	{\bibinfo  {journal} {Journal of Physics G: Nuclear and Particle Physics}\
	}\textbf {\bibinfo {volume} {41}},\ \bibinfo {pages} {035005} (\bibinfo
	{year} {2014})}\BibitemShut {NoStop}%
\bibitem [{\citenamefont {Sikivie}\ \emph {et~al.}(2014)\citenamefont
	{Sikivie}, \citenamefont {Sullivan},\ and\ \citenamefont
	{Tanner}}]{Sikivie2014a}%
\BibitemOpen
\bibfield  {author} {\bibinfo {author} {\bibfnamefont {P.}~\bibnamefont
		{Sikivie}}, \bibinfo {author} {\bibfnamefont {N.}~\bibnamefont {Sullivan}}, \
	and\ \bibinfo {author} {\bibfnamefont {D.~B.}\ \bibnamefont {Tanner}},\
}\href {\doibase 10.1103/PhysRevLett.112.131301} {\bibfield  {journal}
{\bibinfo  {journal} {Phys. Rev. Lett.}\ }\textbf {\bibinfo {volume} {112}},\
\bibinfo {pages} {131301} (\bibinfo {year} {2014})}\BibitemShut {NoStop}%
\bibitem [{\citenamefont {Budker}\ \emph {et~al.}(2014)\citenamefont {Budker},
	\citenamefont {Graham}, \citenamefont {Ledbetter}, \citenamefont
	{Rajendran},\ and\ \citenamefont {Sushkov}}]{BudkerPRX}%
\BibitemOpen
\bibfield  {author} {\bibinfo {author} {\bibfnamefont {D.}~\bibnamefont
		{Budker}}, \bibinfo {author} {\bibfnamefont {P.~W.}\ \bibnamefont {Graham}},
	\bibinfo {author} {\bibfnamefont {M.}~\bibnamefont {Ledbetter}}, \bibinfo
	{author} {\bibfnamefont {S.}~\bibnamefont {Rajendran}}, \ and\ \bibinfo
	{author} {\bibfnamefont {A.~O.}\ \bibnamefont {Sushkov}},\ }\href {\doibase
	10.1103/PhysRevX.4.021030} {\bibfield  {journal} {\bibinfo  {journal} {Phys.
			Rev. X}\ }\textbf {\bibinfo {volume} {4}},\ \bibinfo {pages} {021030}
	(\bibinfo {year} {2014})}\BibitemShut {NoStop}%
\bibitem [{\citenamefont {Arvanitaki}\ and\ \citenamefont
	{Geraci}(2014)}]{altmethods2014}%
\BibitemOpen
\bibfield  {author} {\bibinfo {author} {\bibfnamefont {A.}~\bibnamefont
		{Arvanitaki}}\ and\ \bibinfo {author} {\bibfnamefont {A.~A.}\ \bibnamefont
		{Geraci}},\ }\href {\doibase 10.1103/PhysRevLett.113.161801} {\bibfield
	{journal} {\bibinfo  {journal} {Phys. Rev. Lett.}\ }\textbf {\bibinfo
		{volume} {113}},\ \bibinfo {pages} {161801} (\bibinfo {year}
	{2014})}\BibitemShut {NoStop}%
\bibitem [{\citenamefont {Sikivie}(2014)}]{Sikivie2014}%
\BibitemOpen
\bibfield  {author} {\bibinfo {author} {\bibfnamefont {P.}~\bibnamefont
		{Sikivie}},\ }\href {\doibase 10.1103/PhysRevLett.113.201301} {\bibfield
	{journal} {\bibinfo  {journal} {Phys. Rev. Lett.}\ }\textbf {\bibinfo
		{volume} {113}},\ \bibinfo {pages} {201301} (\bibinfo {year}
	{2014})}\BibitemShut {NoStop}%
\bibitem [{\citenamefont {Kim}(1979)}]{K79}%
\BibitemOpen
\bibfield  {author} {\bibinfo {author} {\bibfnamefont {J.~E.}\ \bibnamefont
		{Kim}},\ }\href {\doibase 10.1103/PhysRevLett.43.103} {\bibfield  {journal}
	{\bibinfo  {journal} {Phys. Rev. Lett.}\ }\textbf {\bibinfo {volume} {43}},\
	\bibinfo {pages} {103} (\bibinfo {year} {1979})}\BibitemShut {NoStop}%
\bibitem [{\citenamefont {Shifman}\ \emph {et~al.}(1980)\citenamefont
	{Shifman}, \citenamefont {Vainshtein},\ and\ \citenamefont
	{Zakharov}}]{SVZ80}%
\BibitemOpen
\bibfield  {author} {\bibinfo {author} {\bibfnamefont {M.}~\bibnamefont
		{Shifman}}, \bibinfo {author} {\bibfnamefont {A.}~\bibnamefont {Vainshtein}},
	\ and\ \bibinfo {author} {\bibfnamefont {V.}~\bibnamefont {Zakharov}},\
}\href {\doibase http://dx.doi.org/10.1016/0550-3213(80)90209-6} {\bibfield
{journal} {\bibinfo  {journal} {Nuclear Physics B}\ }\textbf {\bibinfo
	{volume} {166}},\ \bibinfo {pages} {493 } (\bibinfo {year}
{1980})}\BibitemShut {NoStop}%
\bibitem [{\citenamefont {Dine}\ \emph {et~al.}(1981)\citenamefont {Dine},
	\citenamefont {Fischler},\ and\ \citenamefont {Srednicki}}]{DFS81}%
\BibitemOpen
\bibfield  {author} {\bibinfo {author} {\bibfnamefont {M.}~\bibnamefont
		{Dine}}, \bibinfo {author} {\bibfnamefont {W.}~\bibnamefont {Fischler}}, \
	and\ \bibinfo {author} {\bibfnamefont {M.}~\bibnamefont {Srednicki}},\ }\href
{\doibase http://dx.doi.org/10.1016/0370-2693(81)90590-6} {\bibfield
	{journal} {\bibinfo  {journal} {Physics Letters B}\ }\textbf {\bibinfo
		{volume} {104}},\ \bibinfo {pages} {199 } (\bibinfo {year}
	{1981})}\BibitemShut {NoStop}%
\bibitem [{\citenamefont {Kafle}\ \emph {et~al.}(2014)\citenamefont {Kafle},
	\citenamefont {Sharma}, \citenamefont {Lewis},\ and\ \citenamefont
	{Bland-Hawthorn}}]{cdmdensity2014}%
\BibitemOpen
\bibfield  {author} {\bibinfo {author} {\bibfnamefont {P.~R.}\ \bibnamefont
		{Kafle}}, \bibinfo {author} {\bibfnamefont {S.}~\bibnamefont {Sharma}},
	\bibinfo {author} {\bibfnamefont {G.~F.}\ \bibnamefont {Lewis}}, \ and\
	\bibinfo {author} {\bibfnamefont {J.}~\bibnamefont {Bland-Hawthorn}},\ }\href
{http://stacks.iop.org/0004-637X/794/i=1/a=59} {\bibfield  {journal}
	{\bibinfo  {journal} {The Astrophysical Journal}\ }\textbf {\bibinfo {volume}
		{794}},\ \bibinfo {pages} {59} (\bibinfo {year} {2014})}\BibitemShut
{NoStop}%
\bibitem [{\citenamefont {Walls}(1992)}]{WallsCC}%
\BibitemOpen
\bibfield  {author} {\bibinfo {author} {\bibfnamefont {W.}~\bibnamefont
		{Walls}},\ }in\ \href {\doibase 10.1109/FREQ.1992.270007} {\emph {\bibinfo
		{booktitle} {Frequency Control Symposium, 1992. 46th., Proceedings of the
			1992 IEEE}}}\ (\bibinfo {year} {1992})\ pp.\ \bibinfo {pages} {257
	--261}\BibitemShut {NoStop}%
\bibitem [{\citenamefont {Rubiola}\ and\ \citenamefont
	{Giordano}(2000)}]{Rubiola}%
\BibitemOpen
\bibfield  {author} {\bibinfo {author} {\bibfnamefont {E.}~\bibnamefont
		{Rubiola}}\ and\ \bibinfo {author} {\bibfnamefont {V.}~\bibnamefont
		{Giordano}},\ }\href@noop {} {\bibfield  {journal} {\bibinfo  {journal} {Rev.
			Sci. Instrum.}\ }\textbf {\bibinfo {volume} {71}},\ \bibinfo {pages} {3085}
	(\bibinfo {year} {2000})}\BibitemShut {NoStop}%
\bibitem [{\citenamefont {Rubiola}\ and\ \citenamefont
	{Vernotte}()}]{Rubiola2010}%
\BibitemOpen
\bibfield  {author} {\bibinfo {author} {\bibfnamefont {E.}~\bibnamefont
		{Rubiola}}\ and\ \bibinfo {author} {\bibfnamefont {F.}~\bibnamefont
		{Vernotte}},\ }\href@noop {} {\enquote {\bibinfo {title} {The cross-spectrum
			experimental method},}\ }\Eprint {http://arxiv.org/abs/arXiv:1003.0113v1}
{arXiv:1003.0113v1} \BibitemShut {NoStop}%
\bibitem [{\citenamefont {Darin}(2001)}]{darin2001}%
\BibitemOpen
\bibfield  {author} {\bibinfo {author} {\bibfnamefont {S.}~\bibnamefont
		{Darin}},\ }\emph {\bibinfo {title} {First results from a
		multiple-microwave-cavity search for dark-matter axions}},\ \href@noop {}
{Ph.D. thesis},\ \bibinfo  {school} {UC Davis} (\bibinfo {year}
{2001})\BibitemShut {NoStop}%
\bibitem [{\citenamefont {Shokair}\ \emph {et~al.}(2014)\citenamefont
	{Shokair}, \citenamefont {Root}, \citenamefont {Van~Bibber}, \citenamefont
	{Brubaker}, \citenamefont {Gurevich}, \citenamefont {Cahn}, \citenamefont
	{Lamoreaux}, \citenamefont {Anil}, \citenamefont {Lehnert}, \citenamefont
	{Mitchell}, \citenamefont {Reed},\ and\ \citenamefont {Carosi}}]{ADMXHF2014}%
\BibitemOpen
\bibfield  {author} {\bibinfo {author} {\bibfnamefont {T.~M.}\ \bibnamefont
		{Shokair}}, \bibinfo {author} {\bibfnamefont {J.}~\bibnamefont {Root}},
	\bibinfo {author} {\bibfnamefont {K.~A.}\ \bibnamefont {Van~Bibber}},
	\bibinfo {author} {\bibfnamefont {B.}~\bibnamefont {Brubaker}}, \bibinfo
	{author} {\bibfnamefont {Y.~V.}\ \bibnamefont {Gurevich}}, \bibinfo {author}
	{\bibfnamefont {S.~B.}\ \bibnamefont {Cahn}}, \bibinfo {author}
	{\bibfnamefont {S.~K.}\ \bibnamefont {Lamoreaux}}, \bibinfo {author}
	{\bibfnamefont {M.~A.}\ \bibnamefont {Anil}}, \bibinfo {author}
	{\bibfnamefont {K.~W.}\ \bibnamefont {Lehnert}}, \bibinfo {author}
	{\bibfnamefont {B.~K.}\ \bibnamefont {Mitchell}}, \bibinfo {author}
	{\bibfnamefont {A.}~\bibnamefont {Reed}}, \ and\ \bibinfo {author}
	{\bibfnamefont {G.}~\bibnamefont {Carosi}},\ }\href {\doibase
	10.1142/S0217751X14430040} {\bibfield  {journal} {\bibinfo  {journal}
		{International Journal of Modern Physics A}\ }\textbf {\bibinfo {volume}
		{29}},\ \bibinfo {pages} {1443004} (\bibinfo {year} {2014})}\BibitemShut
{NoStop}%
\bibitem [{\citenamefont {Carois}\ and\ \citenamefont
	{Bibber}(2008)}]{carosi2008}%
\BibitemOpen
\bibfield  {author} {\bibinfo {author} {\bibfnamefont {G.}~\bibnamefont
		{Carois}}\ and\ \bibinfo {author} {\bibfnamefont {K.~V.}\ \bibnamefont
		{Bibber}},\ }in\ \href@noop {} {\emph {\bibinfo {booktitle} {Lecture Notes in
			Physics - Axions: Theory, Cosmology, and Experimental Searches}}},\ \bibinfo
{editor} {edited by\ \bibinfo {editor} {\bibfnamefont {M.}~\bibnamefont
		{Kuster}}, \bibinfo {editor} {\bibfnamefont {G.}~\bibnamefont {Raffelt}}, \
	and\ \bibinfo {editor} {\bibfnamefont {B.}~\bibnamefont
		{Beltr$\acute{a}$n}}}\ (\bibinfo  {publisher} {Springer},\ \bibinfo {year}
{2008})\BibitemShut {NoStop}%
\end{thebibliography}
\end{document}